%% file: main.tex
\documentclass[%
reprint,
superscriptaddress,
showpacs,
nofootinbib,
amsmath,
amssymb,
amsfonts,
aps,
pra,
floatfix,
showkeys
]{revtex4-1}

\RequirePackage{atbegshi}

\input{Header}

\begin{document}

\title{
The light Roberge-Weiss tricritical endpoint at imaginary isospin and baryon chemical potential
}

\author{Bastian B. Brandt}
 \email{brandt@physik.uni-bielefeld.de}
 \affiliation{
  Institute for Theoretical Physics, University of Bielefeld, \\
  D-33615 Bielefeld, Germany
}

\author{Amine Chabane}
 \email{chabane@itp.uni-frankfurt.de}
 \affiliation{
  Institut f\"{u}r Theoretische Physik, Goethe-Universit\"{a}t Frankfurt\\
 Max-von-Laue-Str.\ 1, 60438 Frankfurt am Main, Germany
}

\author{Volodymyr Chelnokov}
 \email{chelnokov@itp.uni-frankfurt.de}
 \affiliation{
  Institut f\"{u}r Theoretische Physik, Goethe-Universit\"{a}t Frankfurt\\
 Max-von-Laue-Str.\ 1, 60438 Frankfurt am Main, Germany
}

\author{Francesca Cuteri}
\email{cuteri@itp.uni-frankfurt.de}
\affiliation{
 Institut f\"{u}r Theoretische Physik, Goethe-Universit\"{a}t Frankfurt\\
 Max-von-Laue-Str.\ 1, 60438 Frankfurt am Main, Germany
}

\author{Gergely Endr\H{o}di}
 \email{endrodi@physik.uni-bielefeld.de}
 \affiliation{
  Institute for Theoretical Physics, University of Bielefeld, \\
  D-33615 Bielefeld, Germany
}

\author{Christopher Winterowd}
\email{winterowd@itp.uni-frankfurt.de}
\affiliation{
 Institut f\"{u}r Theoretische Physik, Goethe-Universit\"{a}t Frankfurt\\
 Max-von-Laue-Str.\ 1, 60438 Frankfurt am Main, Germany
}

\begin{abstract}

Imaginary chemical potentials serve as a useful tool to constrain 
the QCD phase diagram and to gain insight into the thermodynamics of strongly interacting matter. 
In this study, we report on the first determination of the 
phase diagram for arbitrary imaginary baryon and isospin chemical potentials at 
high temperature using one-loop perturbation theory, revealing a nontrivial 
structure of Roberge-Weiss (RW) phase transitions in this plane.
Subsequently, this system is simulated numerically with
$N_f=2$ unimproved staggered quarks on $N_{\tau}=4$ lattices at a range of temperatures at 
one of the RW phase transitions.
We establish a lower bound for the light quark mass, where the first-order transition line terminates in a tricritical point. 
It is found that this tricritical mass is increased as compared to the case of purely baryonic imaginary chemical potentials, 
indicating that our setup is more advantageous for identifying critical behavior towards the chiral limit.
Finally, the dynamics of local Polyakov loop clusters is also studied in conjuction with the RW phase transition.
\end{abstract}

\maketitle

\input{Introduction}
\input{QCDImagQAndI}
\input{LatticeSimulationsAndAnalysis}
\input{LightTricriticalEndpoint}
\input{CenterDomains}
\input{Conclusions}

\acknowledgments
    We thank Alesssandro Sciarra for helpful discussions, especially regarding the quantitative collapse procedure.
    We thank Owe Philipsen for useful discussions and for infrastructural support.
    The authors acknowledge support by the Deutsche Forschungsgemeinschaft (DFG, German Research Foundation) through the CRC-TR 211 ``Strong-interaction matter under extreme conditions''~--~project number 315477589~--~TRR 211. F.C. acknowledges the support by the State of Hesse within the Research Cluster ELEMENTS (Project ID 500/10.006). The authors also gratefully acknowledge the Gauss Centre for Supercomputing e.V. (\href{https://www.gauss-centre.eu}{\tt www.gauss-centre.eu}) for funding this project by providing computing time on the GCS Supercomputer SuperMUC-NG at Leibniz Supercomputing Centre (\href{https://www.lrz.de}{\tt www.lrz.de}). We also thank the computing staff of the \hlr{} cluster for their support.

\bibliographystyle{apsrev4-1}
\bibliography{references}

\onecolumngrid

\appendix
\input{AppendixCarpet}

\input{AppendixSimulations}
\input{AppendixQuantitativeCollapse}

\end{document}

%% file: Header.tex
\usepackage[utf8]{inputenc}
\usepackage[T1]{fontenc}
\usepackage[ngerman,english]{babel}
\usepackage[matha]{mathabx} 
\usepackage[dvipsnames, svgnames, table]{xcolor}
\usepackage{%
    graphicx,
    siunitx,
    multirow,
    booktabs,
    bbold,    
    pifont,   
    capt-of,
    pgfplotstable,
}
\usepackage{subfigure}     
\renewcommand{\thesubfigure}{(\alph{subfigure})}
\makeatletter
  \renewcommand{\@thesubfigure}{\thesubfigure\space}
  \def\@currentlabel{\p@subfigure\thesubfigure}
\makeatother
\usepackage{hyperref}
\hypersetup{
    colorlinks=true,
    citecolor=black,
    linkcolor=black,
    urlcolor=black,
    anchorcolor=black,
    linktocpage
}
\usepackage{etoolbox} 
\usepackage{cleveref} 
\makeatletter
\appto{\appendix}{%
  \@ifstar{\def\theequation@prefix{A.}}%
          {}%
}
\makeatother
\crefname{figure}{Figure}{Figures}
\crefname{table}{Table}{Tables}
\crefname{equation}{Eq.}{Eqs.}
\crefname{section}{Section}{Sections}
\graphicspath{{./figures/}}

\newcommand{\referencename}{Ref.}
\newcommand{\referencesname}{Refs.}
\newcommand{\refcite}[1]{\referencename~[\onlinecite{#1}]}
\newcommand{\refscite}[2]{\referencesname~[\onlinecite{#1}] and~[\onlinecite{#2}]}

\DeclareMathOperator{\Tr}{Tr}

\newcommand{\id}{\mathbb{1}}
\newcommand{\Z}{\mathbb{Z}}
\newcommand{\Nspat}{N_\text{s}}
\newcommand{\Ntau}{N_\tau}

\newcommand{\Nf}{N_\text{f}}
\newcommand{\mLight}{m_\text{ud}}
\newcommand{\muI}{\mu_\text{I}}
\newcommand{\muB}{\mu_\text{B}}
\newcommand{\thetaI}{\theta_\text{I}}
\newcommand{\thetaB}{\theta_\text{B}}
\newcommand{\hlr}{Goethe-HLR}
\newcommand{\Action}{\mathcal S}
\newcommand{\ActionGauge}{\Action_{\text{g}}}

\newcommand{\vev}[1]{ \langle \, #1 \, \rangle }

\newcommand{\Plaq}{U_{\text{p}}}
\newcommand{\betaC}{\beta_\text{c}}

\newcommand{\beq} {\begin{eqnarray}}
\newcommand{\eeq} {\end{eqnarray}}
\newcommand{\nn}{ \nonumber} 

\pgfplotsset{compat=1.17}
\pgfplotstableread[col sep=comma]{
nf,nt,mu,name,mean,errorp,errorm,reference
,,{$(0,\,0)$},Wilson,560,6,6,\cite{Philipsen:2016hkv}
,,{$(\imath\pi,\,0)$},Wilson,913,3,3,\cite{Philipsen:2014rpa}
2,4,{$(0,\,0)$},Staggered,60,12,12,\cite{Bonati:2014kpa}
,,{$(\imath\pi,\,0)$},Staggered,473,29,28,\cite{Bonati:2010gi,Philipsen:2019ouy}
,,{$(\imath\pi, \,\imath\pi/6)$},Staggered,558,56,0, this work
}\data

\newcommand{\errplot}{%
  \begin{tikzpicture}[trim axis left,trim axis right]
    \begin{axis}[
        y=-\baselineskip,
        scale only axis,
        width             = 11cm,
        enlarge y limits  = {abs=0.5},
        axis y line*      = middle,
        y axis line style = dashed,
        ytick             = \empty,
        axis x line*      = bottom
      ]
      \addplot+[mark size=1, only marks, color=blue][error bars/.cd,x dir=both, x explicit]
        table [x=mean,y expr=\coordindex,x error plus=errorp,x error minus=errorm]{\data};
    \end{axis}
  \end{tikzpicture}%
}

%% file: Introduction.tex
\section{Introduction}
Quantum chromodynamics (QCD) has for a long time been established as the fundamental theory governing strong interactions, yet its phase diagram escapes our theoretical understanding in some of its most interesting regions, namely correspondingly to ranges of temperature $T$ and baryon chemical potential $\muB$ that are relevant to the physics of off central heavy-ion collisions, neutron stars’ inner composition and mergers, and, possibly, the early universe around the QCD epoch.
While the inherently non-perturbative nature of strong interactions makes lattice QCD the most suited approach to mapping out the QCD phase diagram, the complex action problem triggered by a nonzero baryon chemical potential represents a major stumbling block to significantly extending in the $\muB$ direction, our current knowledge of the phase diagram at nonzero temperature and vanishing chemical potentials.
The introduction of a purely imaginary baryon chemical potential, for which standard importance sampling techniques are viable, is one of the strategies that have been pursued in order to constrain the phase diagram at real $\muB$ by means of analytic continuation~\cite{deForcrand:2002hgr,DElia:2002tig}.
Moreover, the QCD phase diagram in the plane of temperature and purely imaginary values of the baryon chemical potential has a rich structure that was unveiled long ago~\cite{ROBERGE1986734} and that has been by now the subject of several lattice studies~\cite{DElia:2002tig,deForcrand:2002hgr,deForcrand:2003vyj,deForcrand:2006pv,DElia:2009bzj,Cea:2009ba,deForcrand:2010he,Bonati:2010gi,Bonati:2014kpa,Philipsen:2014rpa,Czaban:2015sas,Philipsen:2016hkv,Bonati:2016pwz, Bonati:2018fvg,Philipsen:2019ouy,Cuteri:2022vwk}.
In the above mentioned physical systems, besides $T$ and $\muB$, there are further thermodynamic parameters affecting the dynamics of strongly interacting matter, namely a nonzero isospin and/or strangeness chemical potential. Despite $\muB$-related effects are typically dominant, a nonzero isospin chemical potential $\muI$, might play a crucial role for the evolution of an early Universe characterised by large (compatibly with observational constraints) lepton flavour asymmetries~\cite{Vovchenko:2020crk,Middeldorf-Wygas:2020glx}.
With all other chemical potentials being zero QCD at nonzero $\muI$ is also a sign-problem-free setup that has been extensively studied over the past few years~\cite{Brandt:2017oyy,Brandt:2018omg}.

Parallel to the attention devoted to the impact on strong interaction matter of the above mentioned thermodynamic parameters, the dependence of the phase diagram on microscopic parameters of the theory, namely the quark masses $m_\text{f}$ has been addressed in many studies in view of the small mass that nature provided the light up and down quarks with and of the `enhanced criticality’ expected on symmetry grounds for the phase diagram of QCD in the chiral limit.
However, due to the singular nature of the fermion determinant at vanishing quark masses, that prevents direct simulations in this limit, extrapolations become necessary in this case too, and it stays an open question\footnote{Though based on recent chirally extrapolated results a first-order phase transition in the continuum for up to six degenerate quark flavors seems to be ruled out~\cite{Cuteri:2021ikv}.} whether in the chiral limit, $m_u=m_d=0$, the chiral phase transition is of first or second order~\cite{Pisarski:1983ms}.
In particular, while results from simulations on coarse lattices, or with unimproved actions, generically indicate that the transition might be of first order, based on results from calculations performed with improved staggered fermions~\cite{Bazavov:2017xul,Bonati:2018fvg,HotQCD:2019xnw,Dini:2021hug} one is rather lead to conclude that the first order nature is, in fact, a lattice artefact.

A further interesting feature of the first order transitions towards the chiral limit is that they have been observed to become significantly stronger as a function of a non-zero, purely imaginary chemical potential, both for staggered \cite{deForcrand:2003vyj,deForcrand:2006pv,Bonati:2014kpa,Bonati:2018fvg} and Wilson \cite{Czaban:2015sas,Philipsen:2016hkv} discretizations.
This observation lead to several attempts to establish the existence of a first order region employing improved fermion discretizations in setups with a nonzero imaginary chemical potential, $\mu\equiv \imath \theta T$.

Along the same lines, in the present manuscript we are testing the effect of a nonzero purely imaginary isospin chemical potential combined with a nonzero purely imaginary baryon chemical potential. First, we map out the corresponding phase diagram at high temperature using both perturbative and non-perturbative approaches. Second, we investigate whether the strength of first order transitions, mass for mass, is further increased by a nonzero purely imaginary isospin chemical potential.
With the purely imaginary baryon chemical potential tuned to take its critical Roberge-Weiss (RW) value (see \cref{sec:imagBAndI}), this is equivalent to establishing the so-called light tricritical mass, by looking at the nature of the transition at the endpoint (the RW endpoint) of the line of first order phase transitions between center sectors \cite{DElia:2009bzj,deForcrand:2010he}, but in the $(T, \imath \muI, \imath \muB)$ space.

Here, we address the nature of the RW endpoint using simulations of $\Nf=2$ unimproved staggered fermions on $\Ntau=4$ lattices in a range for the light quark masses that contains the light tricritical mass as established in analogous simulations carried out at nonzero purely imaginary baryon chemical potential, but at zero purely imaginary isospin chemical potential~\cite{Bonati:2010gi}. We also study the behavior of center domains providing further insight into the Roberge-Weiss phase transition.

Our manuscript is organized as follows: We begin, in \cref{sec:imagBAndI}, with the description of the symmetries and order parameter for the Roberge-Weiss phase transitions in QCD at nonzero imaginary baryon and isospin chemical potential
and the presentation of the perturbative high-temperature phase diagram.
This is followed by, the discussion of our setup for simulations and analysis in \cref{sec:simAndAn}. In \cref{sec:tricrEndpoint} and \cref{sec:centDom} we present and discuss our results on the light tricritical endpoint and on Polyakov loop domains respectively. In three appendices we provide further details on the perturbative calculation, the lattice ensembles and the scaling analysis. 

%% file: QCDImagQAndI.tex
\section{QCD at imaginary baryon and isospin chemical potential}\label{sec:imagBAndI}

\begin{figure}[t]
	\includegraphics[width=\columnwidth]{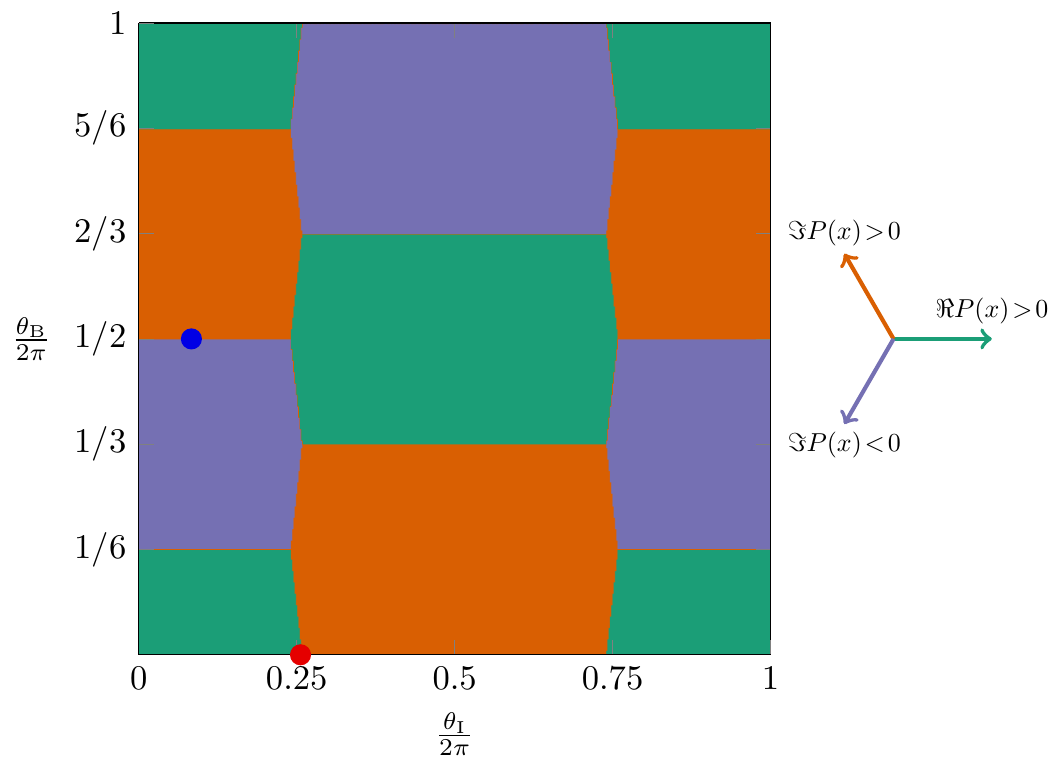}
	\caption{\label{fig:1a} The phase diagram of the theory in the $(\thetaB,\thetaI)$-plane based on the perturbative effective action. The colors denote the orientation of the phase of the volume-averaged Polyakov loop for which the one-loop effective action is minimized. The red dot marks the critical isospin chemical potential at $\thetaB=0$. The blue dot marks the point where we performed dedicated lattice simulations for the RW endpoint. This picture holds at $T\to \infty$ while the situation close to the RW critical point remains an open question. }
\end{figure}

\begin{figure*}
    \includegraphics[width=0.95\linewidth]{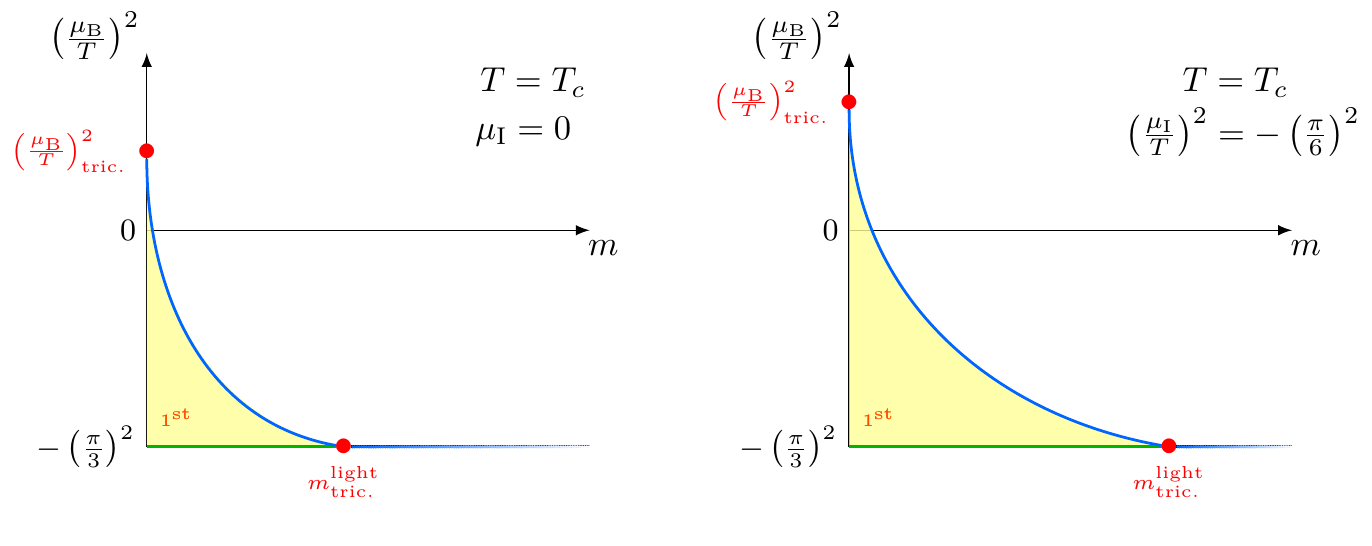}
    \caption{Qualitative sketch of the ``back-plane'' of the three-dimensional $\Nf=2$ Columbia plot in the space of $\mLight$, $\muI$, $\muB$.}
\end{figure*}

QCD in the presence of a purely imaginary baryon potential, $\muB = i \thetaB T$, exhibits a number of symmetries. Firstly, due to charge-conjugation symmetry it is even in $\thetaB$, $Z_{\text{QCD}}(\thetaB) = Z_{\text{QCD}}(-\thetaB)$.
Furthermore, under the combined action of an SU(3) gauge transformation, which satisfies 
\beq
G(\vec{x},\tau+\beta) = H G(\vec{x},\tau), ~ H \in Z(3), 
\eeq
with $\beta \equiv 1/T$, and a shift of the imaginary chemical potential one can show that 
\beq 
Z_{\text{QCD}}(\thetaB) = Z_{\text{QCD}}(\thetaB+2\pi k/3),~k=0,1,2\,.
\eeq 
These symmetries imply a particular phase-structure in the $(T,\thetaB)$-plane. As predicted by perturbative calculations and later confirmed on the lattice, there exist first-order transition lines, oriented parallel to the $T$-axis at critical values of the imaginary baryon chemical potential, $\theta_{k,\text{c}} \equiv (2k+1)\pi/3,~k \in \Z$. These lines terminate at a critical temperature, $T_{\text{RW},\text{c}}$. At low-$T$, the transition becomes a crossover. The sectors, which the first-order lines separate, are characterized by the phase of the expectation value of the Polyakov loop, which is the trace of the Wilson line which winds around the compactified time direction. In the continuum, it takes the form 
\beq \label{eq:continuum_polyakov}
P(\vec{x}) = \frac{1}{3} \Tr \mathcal{P} e^{\imath g\int^{\beta}_0 \mathrm{d}\tau A_0 (\vec{x},\tau)} 
\eeq 
where $g$ is the gauge coupling, $\mathcal{P}$ denotes path-ordering, and $A_{\mu} \equiv A^a_{\mu}\lambda^a$, with $\lambda^a,~a=1,\dots,8$ the generators for SU(3). For pure Yang-Mills theory, the Polyakov loop represents the order parameter for Z(3) center symmetry which is spontaneously broken across the deconfinement transition. With the introduction of fermions, center symmetry is explicitly broken. However, at nonzero imaginary baryon chemical potential, the minimum of the free energy is still classified according to the orientation of the expectation value of the Polyakov loop in the complex plane. Moreover, for certain values of the imaginary baryon and isospin chemical potentials, the explicit symmetry breaking is even absent, see below.

 For $\Nf=2$ mass-degenerate quarks, one can repeat the analysis of Roberge and Weiss, giving each quark flavor a different imaginary potential, say $\theta_{\text{u},\text{d}}$. For this we can introduce the following basis 
\begin{equation} \label{eq:baryon_isospin_chem_potential_basis}
\begin{aligned}
  \theta_{\text{u}} &= \thetaB + \thetaI, \\
  \theta_{\text{d}} &= \thetaB - \thetaI.
\end{aligned}
\end{equation}
One can now ask what are the symmetries of the $\Nf=2$ partition function in terms of $(\thetaB,\thetaI)$. It is immediately clear that the theory is periodic in both variables $\thetaB$ and $\thetaI$, with the period equal to $2\pi$. Furthermore, for two flavors which are degenerate in mass, the theory is also invariant under the transformations 
\begin{align}\label{eq:symmetries_imag_plane} 
    \nn (\thetaB,\thetaI) &\to (\thetaB,-\thetaI), \\ 
    (\thetaB,\thetaI) &\to (\thetaB+\pi,\thetaI-\pi), \\
    \nn (\thetaB,\thetaI) &\to (\thetaB-\pi,\thetaI+\pi). 
\end{align}
The symmetries in~\Cref{eq:symmetries_imag_plane} greatly constrain the shape of the possible phase diagram in the $(\thetaB,\thetaI)$-plane.

At high temperatures, this phase diagram can be calculated via perturbation theory, similarly to the one flavor case~\cite{ROBERGE1986734}.
We discuss the details of this calculation in~\Cref{appendix:RW}, building on~\refscite{AmineBachelorThesis}{Chabane:2021pfk}.
Our final results are conveniently summarized in~\Cref{fig:1a}.
The coloring represents the center sectors of the Polyakov loop, which minimize the one-loop effective potential at different values of $\thetaB$ and $\thetaI$. 
Along the vertical axis, the figure reproduces the well-known structure with first-order phase transitions at $\thetaB/(2\pi) \in \{1/6, 1/2, 5/6\}$.
In turn, along the horizontal axis a transition occurs at the critical value $\thetaI/(2\pi)=\thetaI^{c}/(2\pi)\approx 0.25602409$ marked by the red dot, as well as at $1-\thetaI^{c}/(2\pi)$, in agreement with~\refcite{Cea:2009ba}.
In between these values two phases coexist, implying a Z(2) symmetry that is broken spontaneously (orange and lilac colors in the figure).
Including both imaginary chemical potentials leads to the characteristic ``carpet''-like structure with hexagon-shaped phase regions.
We emphasize that at the meeting point of three hexagons the Z(3) symmetry of the pure gauge theory is recovered (but is, in fact, broken spontaneously).
In~\Cref{appendix:RW}, we also report on the first lattice simulations of this phase diagram that support the validity of the perturbative picture.

%% file: LatticeSimulationsAndAnalysis.tex
\section{Lattice simulations and analysis} \label{sec:simAndAn}

\begin{figure*}
    \centering
   \subfigure[``raw'' (simulated) and   ``rew'' (reweighted) kurtosis]{%
        \label{fig:scaling_plot1}\includegraphics[width=\columnwidth]{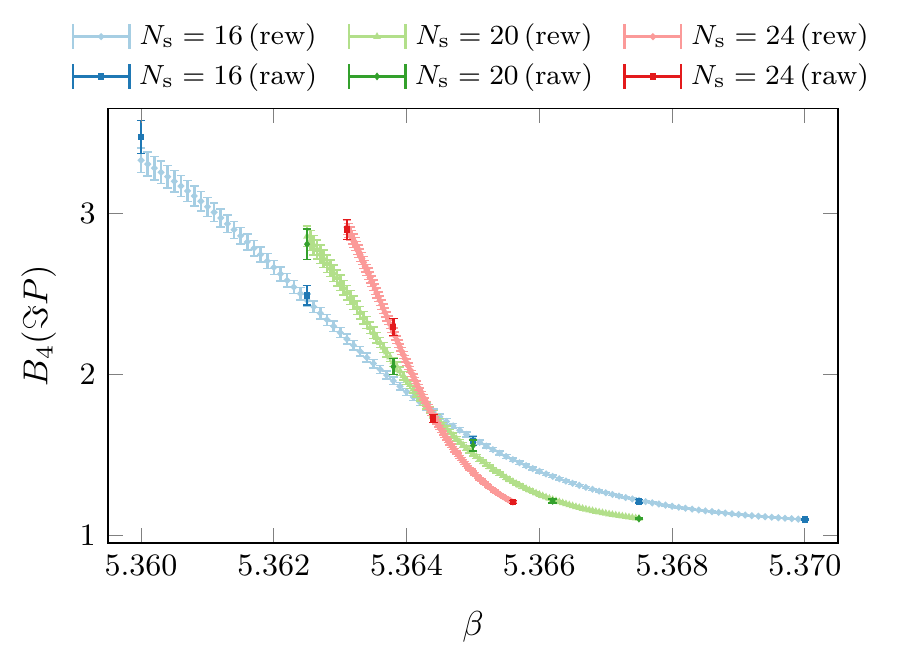}}
   \subfigure[Collapse plot for the reweighted kurtosis]{%
    \label{fig:scaling_plot2}\includegraphics[width=\columnwidth]{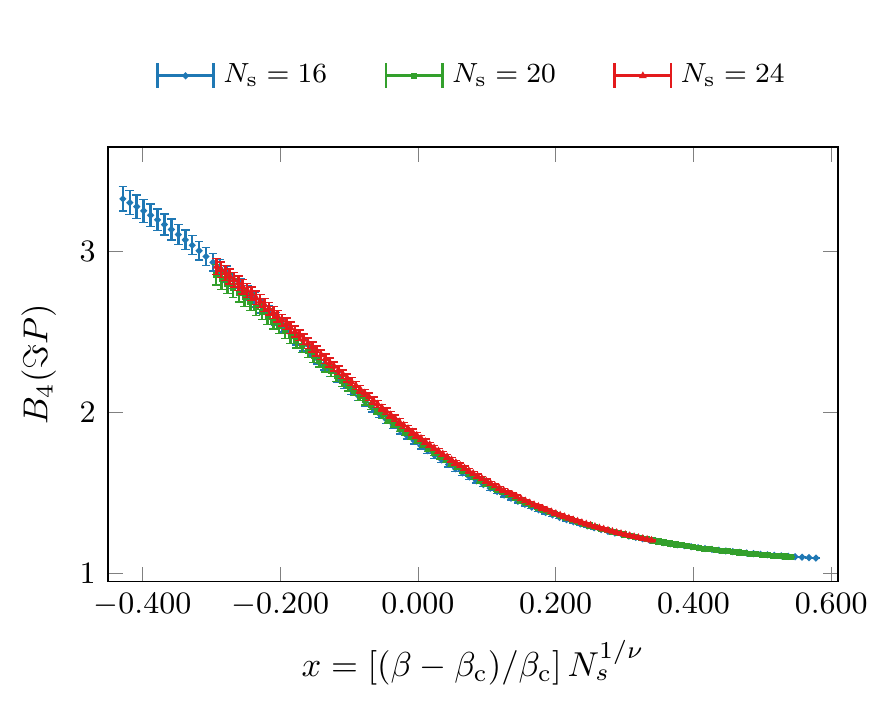}}
  \caption{Results of our analysis for the kurtosis $B_4$ as obtained for quark mass $m=0.04$ and aspect ratios $\Nspat/\Ntau \in \{4,5,6\}$. 
  In~\Cref{fig:scaling_plot1} one notices how the three volumes intersect roughly at the same point.
  The collapse plot in~\Cref{fig:scaling_plot2} shows how all three volumes lie on a single curve around the critical point at $x=0$ and is obtained with the critical parameters $\betaC$ and $\nu$ as obtained via~\Cref{eq:quality_collapse}.}
        \label{fig:scaling_plot}
\end{figure*}

In our simulations we have employed unimproved rooted staggered fermions together with the standard Wilson gauge action.
The rooting of the fermion determinant is needed as each species of staggered fermion in $(3+1)$-dimensions corresponds to four degenerate flavors of Dirac fermions in the continuum limit. 
With regards to improved actions, which aim at reducing lattice discretization errors~\cite{HISQ,ASQTAD_review}, this study is an exploratory one and represents a proof of principle.
The $\Nf=2$ flavors of mass-degenerate staggered fermions at nonzero imaginary isospin and baryon chemmical potential are described by, after the usual Grassmann integration, the following partition function 
\begin{equation} \label{eq:partition_function}
\mathcal{Z}
\!=\!\int \mathcal{D} U e^{-\ActionGauge[U]}
(\det M(\mu_\text{\text{u}}))^{1/4}
(\det M(\mu_\text{\text{d}}))^{1/4},
\end{equation}
where the standard Wilson plaquette action is defined as 
\begin{equation} \label{eq:Wilson_gauge_action}
\ActionGauge = \frac{\beta_{\text{YM}}}{3} \sum_{\text{p}} \operatorname{Re} \Tr_{\text{c}} \left( \id - \Plaq \right), 
\end{equation}
with $\beta_{\text{YM}} \equiv 6/g^2$ and the sum running over all plaquettes of the lattice.
Here, $U_p$ denotes the path-ordered products of gauge links for a given plaquette $p$.
The unimproved staggered Dirac operator reads 
\begin{align} \label{eq:dirac_staggered} \nn
&M(\mu_{\text{f}})_{n,m} \!=\!
\frac{1}{2} \sum^3_{\nu=1} \eta_{n,\nu}\left( U_{\nu}(n) \delta_{n,m\!-\!\hat{\nu}} \!-\!U^{\dagger}_{\nu}(n\!-\!\hat{\nu}) \delta_{n,m\!+\!\hat{\nu}} \right) \\
\nn &\!+  \frac{1}{2}\eta_{n,0}\left( e^{\imath a\theta_{\text{f}} T} U_{0}(n) \delta_{n,m\!-\!\hat{0}} -e^{-\imath a\theta_{\text{f}} T}U^{\dagger}_{0}(n\!-\!\hat{0}) \delta_{n,m+\hat{0}} \right) \\
&\!+ am \delta_{n,m}, 
\end{align}  
where $\eta_{n,\mu}$ are the standard Kawamoto-Smit phases.
The temperature is defined by the temporal extent of the lattice, $T = 1/(a\Ntau)$.
Fixing $\Ntau = 4$, we control the temperature by varying the gauge coupling $\beta_{\text{YM}}$.
Thus, for a fixed bare quark mass, one determines the critical $\beta_{\text{YM},c}$ corresponding to the RW critical point.
A detailed overview of the ensembles generated for this study is presented in~\Cref{appendix:simulations}.

As discussed in the previous section, the primary observable used in determining the RW critical point is the volume-averaged Polyakov loop
\begin{equation} \label{eq:polyakov_loop}
P = \frac{1}{V} \sum_{\vec{x}} \frac{1}{3} \Tr \prod^{N_{\tau}-1}_{\tau=0} U_0(\vec{x}, \tau),
\end{equation}
where the sum runs over all spatial lattice indices and the spatial volume is given by $V \equiv \Nspat^3$.
We simulate at the point $(\pi, \pi/6)$ in the $(\thetaB,\thetaI)$-plane, marked by the light blue dot in~\Cref{fig:1a}.
This point is critical in the $\thetaB$ direction but a safe distance away from the first critical value of $\thetaI$. 
Furthermore, at this point in $\thetaB$ the two Polyakov loop sectors which the first-order line separates at large-$T$ differ in the sign of the imaginary part of $P$.
Thus, as one increases the temperature from the low-$T$ region where the RW transition is a crossover, the distribution of $\operatorname{Im} P$ changes from being a Gaussian centered about zero to being bimodal whose mean is also zero.

The change in the shape of the distribution of $\operatorname{Im} P$ associated with the crossing of the RW critical point can be accurately characterized by studying the normalized moments 
\begin{equation} \label{eq:norm_moments}
B_n(\beta, m, \theta_i) \equiv \frac{\vev{\left(\mathcal{O} - \vev{\mathcal{O}}\right)^n}}{\vev{(\mathcal{O} - \vev{\mathcal{O}})^2}^{n/2}},
\end{equation} 
where $\mathcal{O}$ represents a generic lattice observable. For the case of $\mathcal{O} = \operatorname{Im} P$, it is clear that Z(3) center symmetry demands $\vev{\operatorname{Im} P} = 0$. 
We can enforce this by hand knowing that a finite ensemble generated from a Markov chain will have a nonzero mean with some associated statistical error.
The moment $B_4$, commonly referred to as the kurtosis, will be used in our analysis. 
In particular, one expects its value to vary from $3$ in the low-$T$ limit to $1$ in the high-$T$ limit.
The low-$T$ behavior, however, has been found to be subject to finite-volume effects~\cite{Czaban:2015sas}.

To accurately determine the order of the phase transition, a finite-volume scaling analysis will be performed on the kurtosis.
This will allow us to determine $\betaC$, as well as the critical exponent $\nu$ and the value of the kurtosis at the critical point, $B_{4,\text{c}}$, which are universal in the thermodynamic limit, $V\to \infty$, and dependent on the nature of the phase transition.
As the kurtosis is dimensionless as defined in~\Cref{eq:norm_moments}, the scaling form is written as follows
\begin{equation} \label{eq:kurtosis_scaling}
B_4(\beta; N_s) = g(x) \equiv  g((\beta-\betaC) N^{1/\nu}_s),
\end{equation} 
where $g$ is a universal scaling function and we have also introduced the scaling variable $x \equiv (\beta-\betaC) N^{1/\nu}_s$. What~\Cref{eq:kurtosis_scaling} tells us is that as we approach the thermodynamic limit, the kurtosis computed for $N_v$ different spatial volumes and a fixed quark mass should collapse when plotted as a function of the scaling variable $x$. An efficient and generic way to determine the critical parameters from the scaling behavior of the kurtosis is to define a quantity
\begin{align} \nn
Q\left(\beta, \nu; \{N^{(i)}_s\}\right) &= \frac{1}{2\Delta x} \int^{+\Delta x}_{-\Delta x} \mathrm{d}x \bigg[ N_V \sum_i \left(B_4(x)\right)^2 \\ \label{eq:quality_collapse} & - \left( \sum_i B_4(x) \right)^2 \bigg],
\end{align}
which we refer to as the quality of the collapse.
The quantity $Q$ estimates the average variance of the kurtosis and is minimized as a function of both $\betaC$ and $\nu$.
Here, the interval over which we integrate, $[ -\Delta x, +\Delta x]$, is symmetric about the origin with $\Delta x$ being chosen appropriately such that a final estimate for the critical parameters can be obtained in the limit $\Delta x \to 0$. 

This collapse optimization can be performed including different sets of aspect ratios for one mass.
For $a m = 0.04$ the intersection point for the kurtosis plots at $\Nspat/\Ntau \in \{3, 4\}$ is visibly different than the common one for $\Nspat/\Ntau \in \{4, 5, 6\}$ (5.36490(17) {\it vs} 5.36435(5)).
The results for the latter set of aspect ratios are shown in~\Cref{fig:scaling_plot1}.
That means that the aspect ratio $3$ at this mass is too far from the thermodynamic limit for the kurtosis scaling in~\Cref{eq:kurtosis_scaling} to work. 
Optimization for the sets $\Nspat/\Ntau = \{4,5\}$, $\{5, 6\}$ and $\{4, 5, 6\}$, the latter optimization is shown in~\Cref{fig:scaling_plot2}, results in $\nu$ values $0.46(4)$, $0.411(31)$ and $0.435(17)$ respectively, that are within error bars from each other. 
Still we see that the value of $\nu$ decreases when introducing larger aspect ratio into optimization, which is reasonable since we expect the $\nu$ value in the thermodynamic limit to be $1/3$ for the first order transition region. 
For our final result we used the collapse optimization of $\Nspat/\Ntau = \{4,5\}$ for all three masses we studied.

Further details related to the quality of the collapse can be found in~\Cref{appendix:collapse_details}.
There, the multi-step procedure which was used to obtain the final values of the critical parameters is described.
The consistency of this approach gives one confidence in the measured values and their estimated errors for the aspect ratios considered in this study.  

%% file: LightTricriticalEndpoint.tex
\section{Results on the light Roberge-Weiss tricritical endpoint} \label{sec:tricrEndpoint}

\begin{figure*}
\pgfplotstablegetrowsof{\data}
\let\numberofrows=\pgfplotsretval

\pgfplotstabletypeset[columns={nf,nt,mu,name,errorp,reference},
  every head row/.style = {before row=\toprule, after row=\midrule},
  every last row/.style = {after row=[3ex]\bottomrule},
  after row ={[0ex]},
  columns/nf/.style = {string type, column name={$\Nf$}},
  columns/nt/.style = {string type, column name={$\Ntau$}},
  columns/mu/.style = {string type, column name={$(\thetaB, \,\thetaI)$}},
  columns/name/.style = {string type, column name=Action},
  columns/reference/.style = {string type, column name={Ref.}},
  columns/errorp/.style = {
    column name = {$m_{\pi,\text{light}}^{\text{Z}_2/\text{tricr.}}$ [MeV]},
    assign cell content/.code = {
    \ifnum\pgfplotstablerow=0
    \pgfkeyssetvalue{/pgfplots/table/@cell content}
    {\multirow{\numberofrows}{11.5cm}{\errplot}}%
    \else
    \pgfkeyssetvalue{/pgfplots/table/@cell content}{}%
    \fi
    }
  }
]{\data}
    \caption{Overview of light (tri)critical values of the pion mass from the present and previous studies. The asymmetric 10\% error on the result of the present work is only meant to indicate that our estimate of $m_{\pi,\text{light}}^{\text{tricr.}}=550$~MeV is to be taken a lower bound obtained by using $m=0.06$ for our bare tricritical mass, and the result for the pion mass at the tricritical point from~\refscite{Bonati:2010gi}{Philipsen:2019ouy}.}
    \label{fig:compare}
\end{figure*}

\begin{figure}
	\centering
	\label{fig:sub1}\includegraphics[width=\columnwidth]{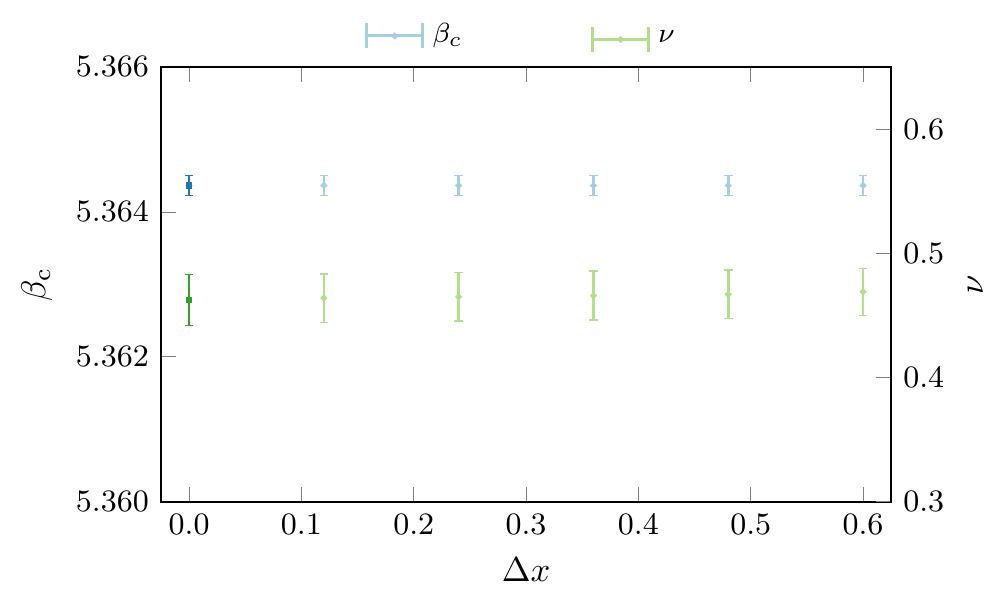}
    \caption{\label{fig:critical_extrap}The dependence of the critical parameters, obtained from the minimization of $Q$, as a function of $\Delta x$ for $am=0.04$ using $\Nspat\in\{16,20\}$. The errors on each each point were estimated using $500$ bootstrap estimators for the kurtosis. The result (shaded points) is clearly stable with respect to $\Delta x$, and thus one can safely extrapolate to $\Delta x=0$ (solid points).}
\end{figure}

\begin{figure} 
    \includegraphics[width=\columnwidth]{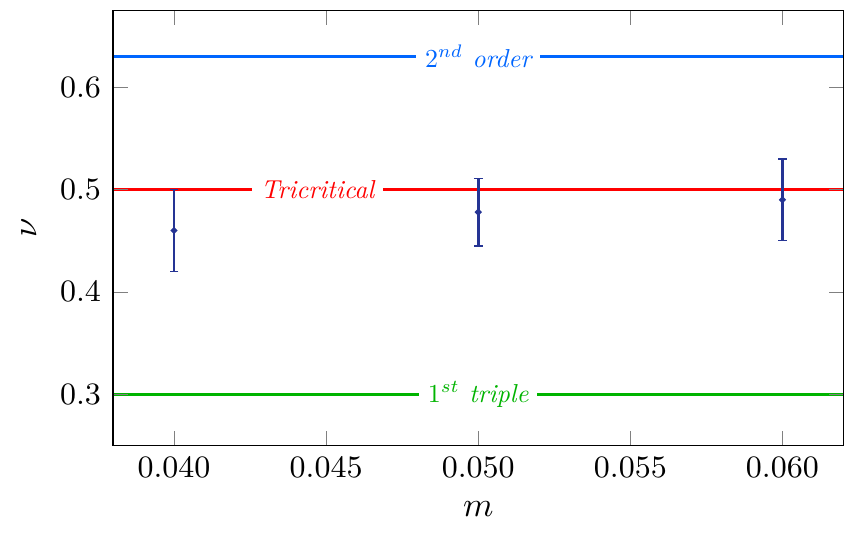}
    \caption{The final, extrapolated to $\Delta x \to 0$, estimates for $\nu$ as a function of quark mass. Each of these points was determined using aspect ratios $4$ and $5$. From this, one can estimate a bound on the lower tricritical mass.}
    \label{fig:final_nu_vs_m}
\end{figure}

An accurate determination of the critical properties of the theory defined by the partition function in~\cref{eq:partition_function} in the RW plane has implications for establishing the order of the transition in the chiral limit at $\mu=0$ as well as for the QCD phase diagram in the $(T,\muB)$ plane.
While not to be expected \textit{a priori}, it was empirically observed that the RW and chiral phase transition temperatures coincide~\cite{Cuteri:2022vwk} in the chiral limit and that the width of the first order triple region in $\Nf=2$ QCD at $\muB=\imath\mu^\text{crit}_\text{RW}$ is much larger than that of the first order region at $\mu=0$ (see~\cref{fig:compare}) making the former significantly less expensive to be reached than the latter.

With knowledge of the tricritical point, in previous studies the $\text{Z}_2$ line stemming from the light RW tricritical point in the $(m_{\text{u,d}}, \muB)$ plane was mapped-out and then extrapolated towards the target tricritical point expected in the chiral limit, in order to determine the order of the chiral phase transition at $\muB=0$.
To achieve this, one is interested in determining the nature of the transition at the end point, $\betaC$, as a function of the quark mass.
At very small and very large quark mass, the RW endpoint is known to be a first-order triple point.
At intermediate masses the RW endpoint becomes a second-order critical point.
This critical point is known to be in the $3D$ Ising ($\text{Z}_2$) universality class.
In this situation, the first-order RW line is analogous to the first-order line lying along the $T$-axis for the Ising model with the endpoint corresponding to $T_c$ and the deviation in the $\thetaB$-direction corresponding to the external magnetic field.
There exist two tricritical masses $m_{\text{tc}}$, which separate the $\text{Z}_2$ regions from the first-order triple point regions.
Previous studies with both staggered and Wilson fermions have obtained estimates for the lower tricritical mass $m_{\text{ltc}}$ \cite{Czaban:2015sas,PhilipsenSciarra2020}. 

It is unknown how the tricritical masses are affected when one turns on a nonzero imaginary isospin chemical potential.
These values are not restricted by any symmetry and thus can shift to either larger or smaller values.
Considering the value $am_{\text{ltc}}=0.043(5)$ of the bare quark mass corresponding to the light tricritical RW endpoint as obtained in previous studies~\cite{Bonati:2010gi} with the same fermion discretization and $\Ntau$, but with $\thetaI=0$, we have performed simulations in the range $am \in [0.04, 0.06]$.
Using finite-size scaling methods for the kurtosis, described in~\cref{sec:tricrEndpoint}, we were able to obtain the critical exponent $\nu$, as a function of the bare quark mass.
As the universal values for $\nu$ in each scenario are known, this allows us to estimate a lower bound for the lower tricritical mass at nozero imaginary isospin.
From the results depicted in~\cref{fig:final_nu_vs_m}, one can estimate that $am_{\text{ltc}} \geq 0.06$.
This result indicates that imaginary isospin chemical potential shifts the bare light tricritical mass to even larger values.
This is also shown in~\cref{fig:compare}.

%% file: CenterDomains.tex
\section{Results on Center Domains} \label{sec:centDom}

\begin{figure*}[t]
   \centering
   \subfigure[]{%
	\label{fig:local_polyakov_absolute}
	\includegraphics[width=\columnwidth]{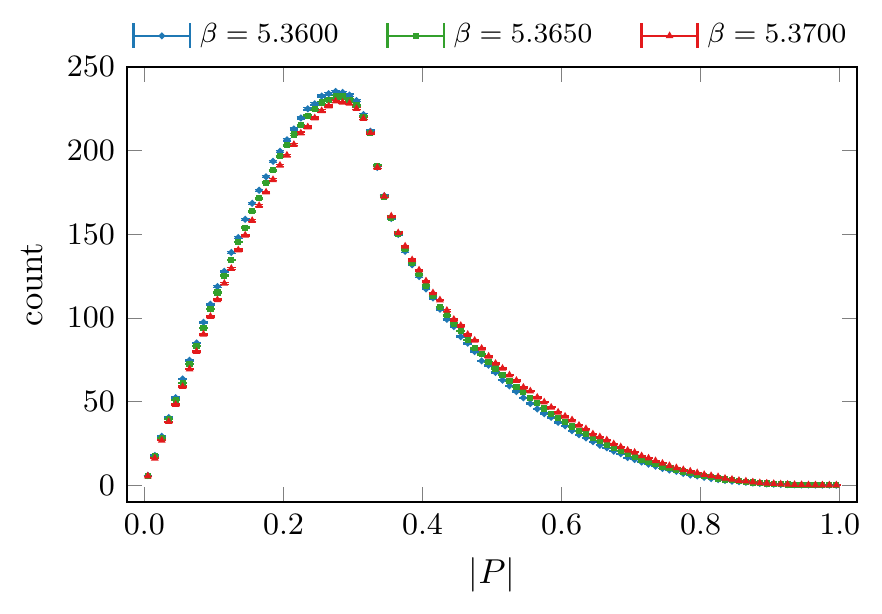}
   }
   \subfigure[]{%
	\label{fig:local_polyakov_angular}
	\includegraphics[width=\columnwidth]{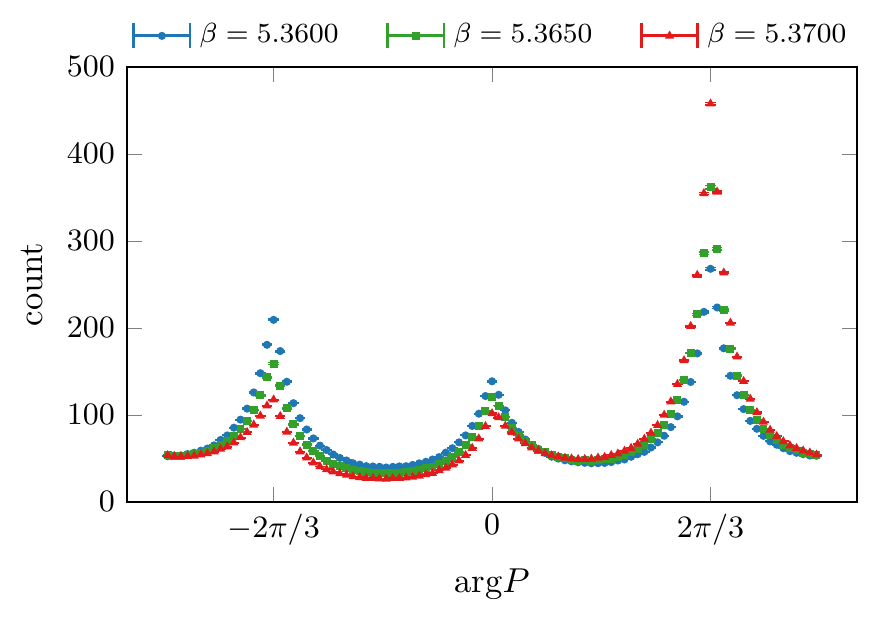}
   }
\caption{\label{fig:local_polyakov}The dependence of the average number of Polyakov loops per configuration having a given absolute value (a), or given phase (b) for $m=0.0400$, $\Nspat = 20$. The data points are connected by straight lines to better visualise the shape of the distributions.}
\end{figure*}

\begin{figure} 
    \includegraphics[width=\columnwidth]{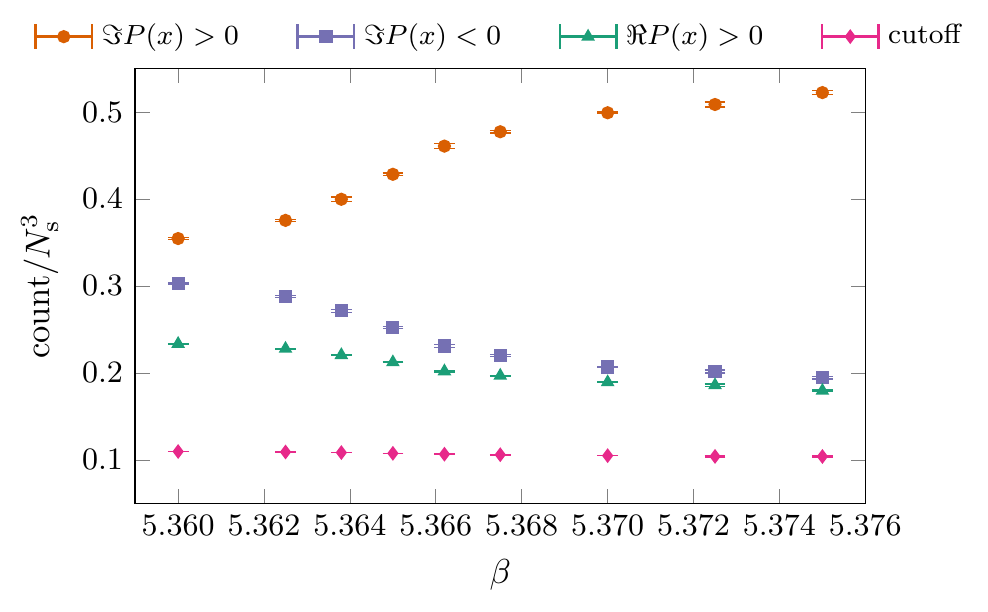}
    \caption{The average number of local Polyakov loops belonging to each of the sectors defined in~\cref{eq:sector_definition} as a function of $\beta$, for $m=0.04$, $\Nspat=20$, $\delta=0.2$.}
    \label{fig:abundances}
\end{figure}

\begin{figure*}[tb]
   \centering
   \subfigure[$\beta=5.3600 < \betaC$]{%
	\label{fig:clusters_small_beta}
	\includegraphics[width=\columnwidth]{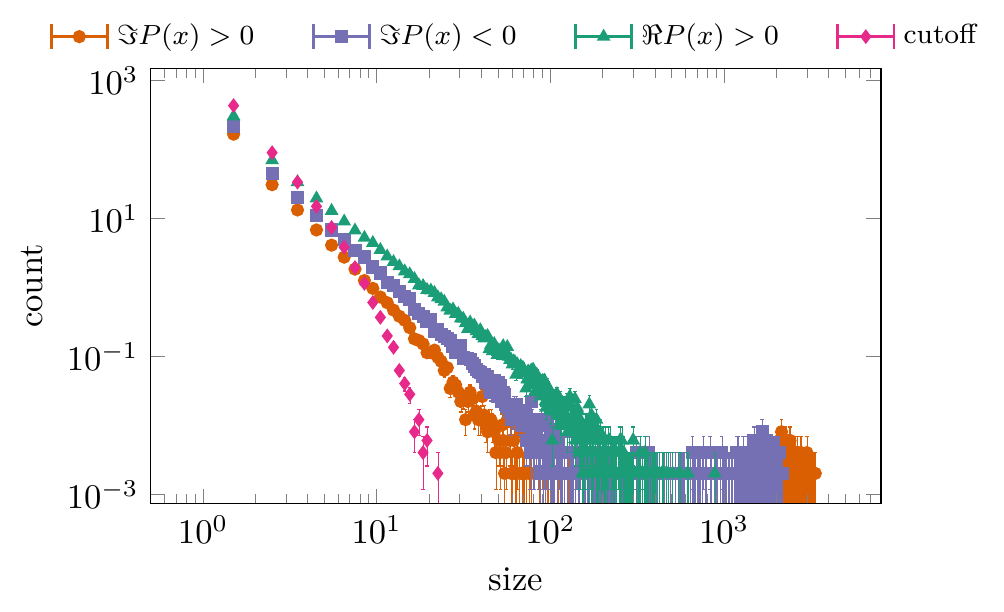}
   }
   \subfigure[$\beta=5.3700 > \betaC$]{%
	\label{fig:clusters_large_beta}
	\includegraphics[width=\columnwidth]{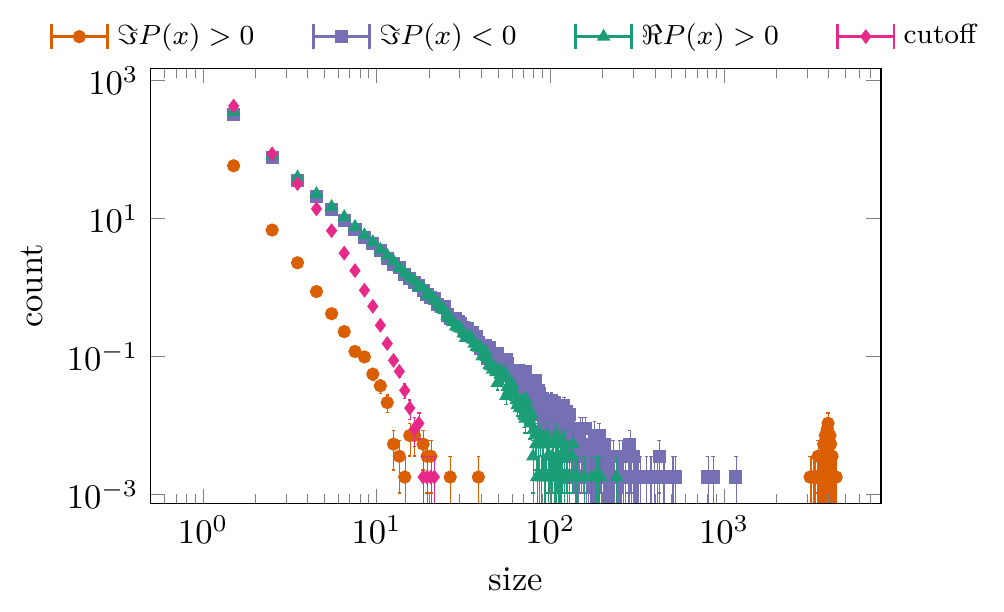}
   }
\caption{\label{fig:clusters} The average number of connected clusters of a given size for $m=0.04$, $\Nspat = 20$ at small and large $\beta$.}
\end{figure*}

Further insight into the properties of the Roberge-Weiss transition can be gained if one studies not just the volume-averaged Polyakov loop but also its local distribution, i.e.\ the Polyakov loop domains. Moreover, local Polyakov loops are intimately connected to localized eigenmodes of the Dirac operator, which have also been studied at the RW transition~\cite{Cardinali:2021fpu}. Polyakov loop domains were previously used to study the deconfinement phase transition both in pure Yang-Mills theories \cite{Fortunato2003,GATTRINGER2010179,EndrodiGattringer2014,Ivanytskyi:2016wcp}, and even in QCD with dynamical fermions (despite the Polyakov loop not strictly being an order parameter of full QCD) \cite{Danzer2011}. Since the Polyakov loop is an order parameter of the Roberge-Weiss phase transition, we can expect this approach to be as useful in our case as in Yang-Mills theory.

In the symmetric phase the volume-averaged Polyakov loop is fluctuating around zero, 
while in the broken symmetry phase the phase space is divided into two symmetric regions 
fluctuating around finite positive and negative values of the imaginary part of the Polyakov loop.
To study the symmetry breaking that occurs at $\beta > \betaC$, we consider in further calculations a subset of generated configurations belonging to one of the sectors,  namely the one in which the imaginary part of the volume-averaged Polyakov loop $P$ is positive.
On each such configuration, we define a local Polyakov loop value $P(x)$, and  decompose it into its absolute value 
$\rho(x)$ and phase $\phi(x)$, 
\begin{equation} 
P(x) = \frac{1}{3} \Tr \prod_{\tau = 0}^{N_t - 1} U_0(x, \tau) = \rho(x) e^{\imath \phi(x)} \ .
\end{equation}

While the distribution of the absolute value of the local Polyakov loop does not change much with $\beta$, the distribution of the phase becomes more and more asymmetric as $\beta$ increases. This is clearly displayed in~\cref{fig:local_polyakov}. To get a numeric estimate of this asymmetry, we assign each spatial site to a sector, according to the following rule~\cite{GATTRINGER2010179}: 
\begin{align}
 \Im P(x) > 0,\  & \text{for $\phi \in  [ \frac{\pi}{3}+\delta, \pi-\delta ]$} \nonumber \\
 \Im P(x) < 0,\  & \text{for $\phi \in  [ -\pi+\delta, -\frac{\pi}{3}-\delta ]$} \nonumber \\  
 \Re P(x) > 0,\  & \text{for $\phi \in  [ -\frac{\pi}{3}+\delta, \frac{\pi}{3}-\delta ]$} \nonumber \\  
 \text{vacuum},\   & \text{otherwise}
 \label{eq:sector_definition}
\end{align}
where $\delta$ is a cutoff parameter that can be freely tuned, and sector 0 is denoted as the ``cutoff'' sector. 

Looking at the dependence of the average population of each sector on $\beta$, shown in~\cref{fig:abundances}, one can see that the population of the sector with $\Im P(x) > 0$ increases with $\beta$, while the population of the sector with $\Im P(x) < 0$ decreases, becoming close to the population of the sector with $\Re P(x) > 0$ at large $\beta$. 
The figure also shows that at each value of $\beta$, the populations of the sectors with $\Im P(x) > 0$ and $\Im P(x) < 0$ are clearly different. 
This illustrates the fact that while the behaviour of the sector abundances is an indication of the phase transition, it does not allow one to resolve the exact location of the transition in our region ($\beta/\betaC = 1 \pm 10^{-3}$).

Another interesting observable which distinguishes the symmetric from the asymmetric phase is the distribution of the sizes of clusters - connected components of spatial sites belonging to the same sector. 
Formally integrating over the fermion fields and the space-like gauge fields one can formulate QCD in terms of local Polyakov loops - the only remaining gauge-invariant degrees of freedom. Considering that the local Polyakov loops are concentrated around the $Z(3)$ center values, the theory can be approximated with a generalized Potts model, where the partition sum can be defined as a sum over the sets of sizes of clusters of spins aligned in the same direction. 
One can extract the expected distribution of cluster sizes from such approximated model and compare with the cluster size distribution obtained from a full theory simulation, setting the model parameters, and obtaining a check of the model overall validity as a description of full theory. 
An example of such a check is done for in pure gauge SU(2) theory in \cite{Ivanytskyi:2016wcp}, which showed good agreement with the liquid droplet model \cite{PhysicsPhysiqueFizika.3.255}. 
We can expect that a similar strategy would also work in our case. 
For this work, instead of comparing the distributions with a fixed model, we just use the fact that a coexistence of two phases at a phase transition point is a sign of a first-order phase transition. 
For an effective Potts cluster model that would mean the existence of ``large'' clusters (clusters with size proportional to the lattice size) in both sectors with $\Im P(x) > 0$ and $\Im P(x) < 0$.

The cluster size distributions for QCD at imaginary values of the baryon and isospin chemical potential are shown in~\cref{fig:clusters}. One can see that for the sector with $\Im P(x) > 0$, this distribution consists of one largest cluster, whose size is proportional to the spatial size of the lattice (about $\Nspat^3 / 2$), and a ``sea'' of smaller ones.
For $\beta > \betaC$, the distribution in other sectors contains just the ``sea'', though the size of these small clusters is on average larger than the size of the ``sea'' clusters in the sector with $\Im P(x) > 0$.
This is due to the fact that in the sector with $\Im P(x) > 0$ the dominant cluster adsorbs a larger part of the ``sea''. 
In this case, the distribution of the sizes in the ``sea'' for the sectors with $\Im P(x) < 0$ and $\Re P(x) > 0$ is similar. 
On the other hand, 
for $\beta < \betaC$, the dominant cluster in the sector with $\Im P(x) > 0$  becomes smaller,
and a smaller cluster in the sectors with $\Im P(x) < 0$ with size proportional to the lattice size can form. 
Here the ``sea'' size distribution has a clear hierarchy: $n_{\Re P(x) > 0}(s) > n_{\Im P(x) < 0}(s) > n_{\Im P(x) > 0}(s)$ for small cluster size $s$,
which can be explained by adsorption of ``sea'' clusters into dominant clusters in both sectors with nonzero imaginary Polyakov loop component. This behavior is clearly visible in~\cref{fig:clusters} and is similar to the one shown for SU(2) Yang-Mills theory in \cite{Ivanytskyi:2016wcp}. 
The difference between the two types of distributions is clearly visible even at this small distance from the transition point. 

%% file: Conclusions.tex
\section{Conclusions} \label{sec:concl}

In this paper we studied $N_f=2$-flavor QCD at nonzero imaginary quark and isospin chemical potentials. Generalizing the calculation of Roberge and Weiss~\cite{ROBERGE1986734}, we calculated 
the associated phase diagram at high temperatures perturbatively and cross-checked it with first lattice simulations. We proceeded by selecting a particular point on this phase diagram that corresponds to a first-order phase transition at high $T$ and performed a dedicated scaling 
study using staggered fermions. The nature of the transition at the RW endpoint as a function of quark mass could be determined by studying the distribution of the imaginary part of the Polyakov loop. Carrying out a finite-size scaling analysis on the kurtosis allowed an identification of a lower bound in quark mass for the light tricritical mass, $m_{\text{tc}} \geq 0.06$. This was aided by the development of an accurate and reproducible method for obtaining both the critical coupling $\betaC$ and the critical exponent $\nu$ through a quantitative fit collapse which does not rely on fit-Ans\"atze and can accurately estimate the associated errors. An exploratory study of the dynamics of local Polyakov loop domains has shown that the distribution of the absolute value of the local Polyakov loops does not change significantly across the transition, while the distribution of the phases of the local Polyakov loops becomes increasingly asymmetric when  $\beta$ increases.

The results we obtained in our proof of concept simulations at nonzero imaginary quark and isospin chemical potential show that the addition of imaginary isospin shifts the light tricritical mass to larger values. This result indicates that our setup is so far the most advantageous for mapping out the critical $\text{Z}_2$ line as it bends towards smaller masses while the values of the imaginary chemical potentials are reduced, and for searching for first order phase transitions in the chiral limit employing improved discretizations, with which so far only upper bounds for the critical/tricritical mass values could be established.

%% file: AppendixCarpet.tex
\section{Effective potential and phase diagram at high temperature}\label{appendix:RW}

In this appendix we discuss the one-loop effective potential at nonzero imaginary baryon and isospin chemical potentials. 
The calculation is based on the perturbative analysis due to Roberge and Weiss~\cite{Weiss:1980rj,Weiss:1981ev,ROBERGE1986734}.
Some of the details have already been presented in~\refscite{AmineBachelorThesis}{Chabane:2021pfk}.

We treat the Polyakov loop in terms of a homogeneous background SU(3) color field $A_0$ and investigate the effective potential for it as a function of the chemical potentials.
After diagonalizing the color field, the Polyakov loop $P$ is parameterized via a sum over the exponentialized eigenvalues
\begin{equation}
 P= \frac{1}{3}\left(e^{\imath \phi_1}+e^{\imath \phi_2}+e^{-\imath \phi_3}\right)\,, \qquad \phi_3=-\phi_1-\phi_2\,,
\end{equation}
where $0\le \phi_{1,2}< 2\pi$.
The constraint for the third eigenvalue ensures that the untraced Polyakov loop has unit determinant.

The effective potential $V_{\rm eff}$ is obtained as the sum of a gluonic $V_{\rm eff}^{G}$ and a fermionic contribution $V_{\rm eff}^F$.
For each quark flavor $\text{f}$, the chemical potential shifts the color field as $A_0\to A_0+\theta_{\text{f}} \mathbb{1}$ and therefore the eigenvalues as $\phi_{j}\to\phi_{j}+\theta_{\text{f}}$. 
Generalizing the results of~\refscite{Weiss:1980rj}{Weiss:1981ev} to two (massless) flavors, we obtain
\begin{align}
   V_{\rm eff}^G(\phi_1,\phi_2) &= \frac{\pi T^{4}}{24} \sum^{3}_{j,k=1} \left\{ 1-\left[\left(\frac{\phi_{j}}{\pi} -\frac{\phi_{k} }{\pi}\right)_{\!\rm mod \;2} -1\right]^{2} \right\}^{2}\,, \\ \label{eff_ferm}
   V_{\rm eff}^F (\phi_1, \phi_2; \theta_{\text{u}},\theta_{\text{d}}) &= -\frac{\pi T^{4}}{12} \sum_{\text{f}=\text{u},\,\text{d}} \sum^{3}_{j=1} \left\{ 1-\left[\left(\frac{\phi_{j}+\theta_{\text{f}}}{\pi} + 1\right)_{\!\rm mod \;2} -1\right]^{2} \right\}^{2}.
\end{align}
Note that $V_{\rm eff}^F$ depends explicitly on the chemical potentials, while $V_{\rm eff}^G$ is independent of them.

\begin{figure}[h]
    \centering
    \includegraphics[scale=0.46]{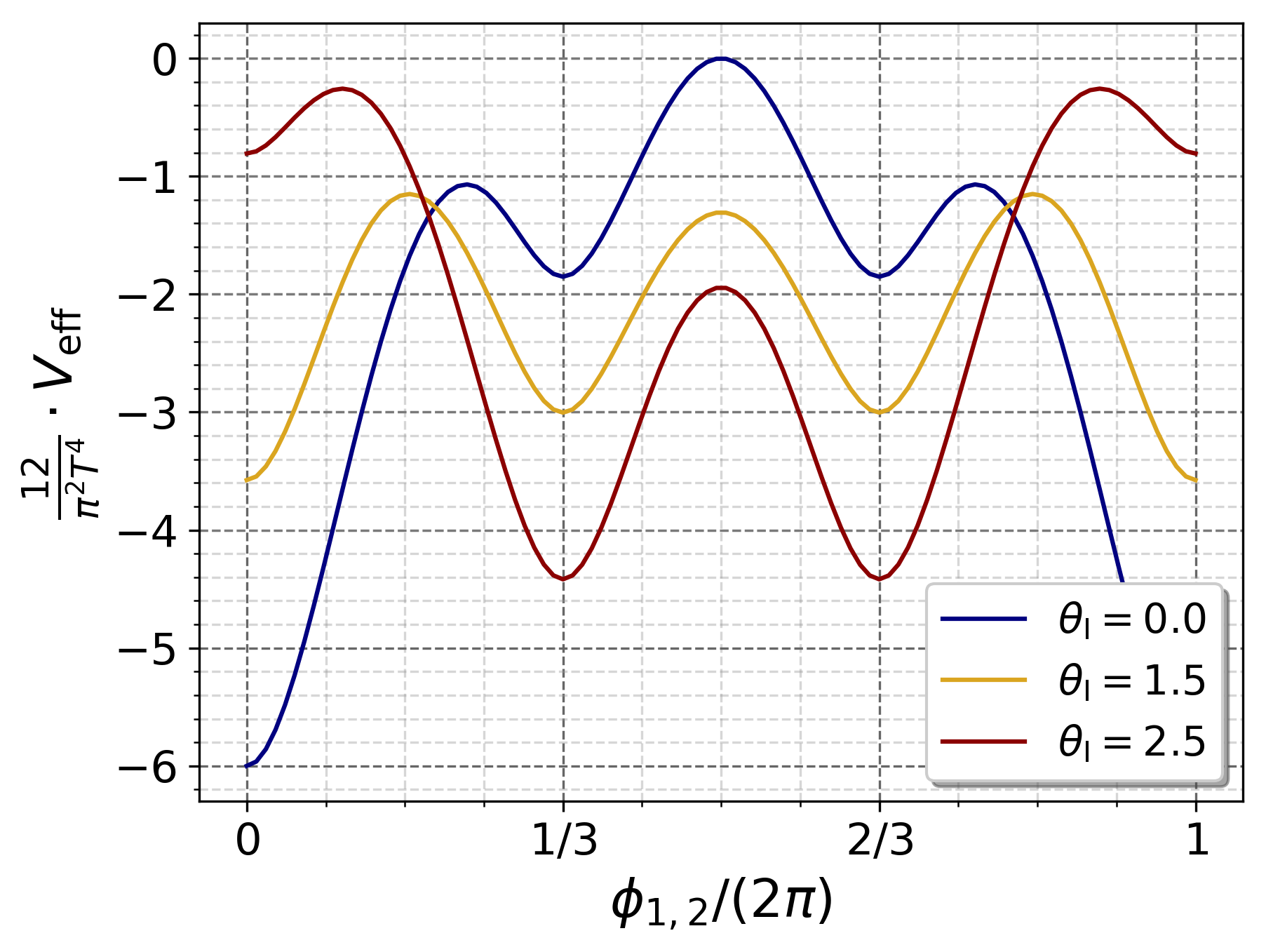}\qquad
    \includegraphics[scale=0.46]{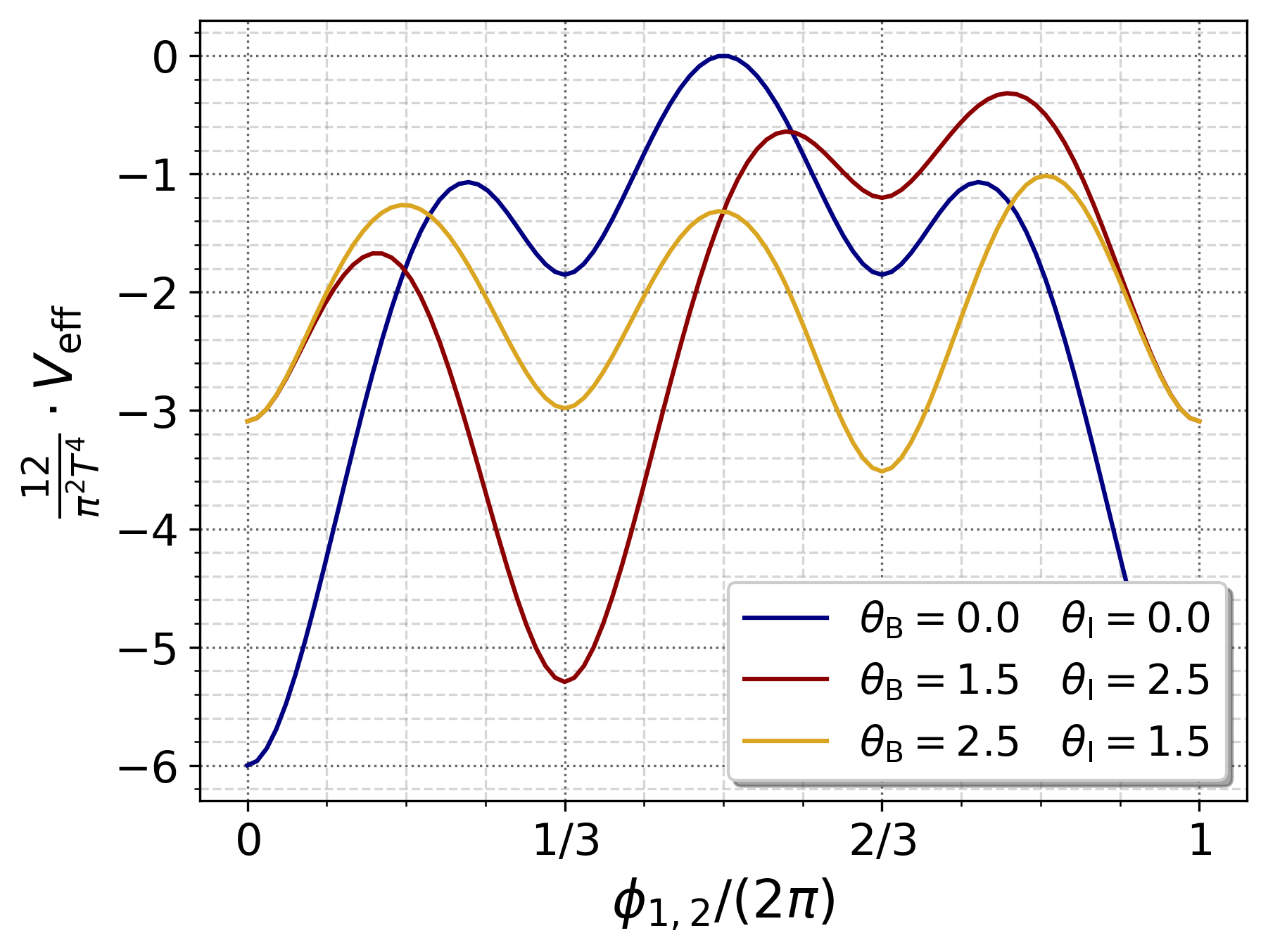}
    \caption{The normalized effective potential as a function of the fields $\phi_{1,2}/(2\pi)$ for different values of $\thetaB$ and of $\thetaI$. In each case the minimum is located at one of the 
    center sectors.}
    \label{fig:eff_iso}
\end{figure}

Identifying the ground state as a function of $\theta_{\text{u}}$ and $\theta_{\text{d}}$ -- or, equivalently, as a function of $\thetaB$ and $\thetaI$ -- amounts to minimizing $V_{\rm eff}$ in $\phi_{1,2}$. 
Just like in the case without isospin chemical potential~\cite{ROBERGE1986734}, the minima always occur at $\phi_1=\phi_2=\{0,2\pi/3, -2\pi/3\}$, corresponding to the three center sectors $P=\{1$, $e^{\imath 2\pi/3}$,
$e^{-\imath 2\pi/3}\}$ of the Polyakov loop.
This is exemplified in~\cref{fig:eff_iso}, which shows the effective potential as a function of $\phi_1=\phi_2$ for various values of both chemical potentials.
Mapping the complete range of the parameters $\thetaB$ and $\thetaI$ gives the phase diagram shown in~\cref{fig:1a}. 

To cross-check the perturbative results, we also considered full lattice QCD simulations of the same phase diagram. 
We use the action discussed in~\cref{sec:simAndAn} with an inverse gauge coupling of $\beta_{\rm YM}=5.2$ and degenerate quark masses $ma=0.025$.
This coincides with the setup of~\refcite{Endrodi:2014lja}, where real isospin chemical potentials were considered.
We perform simulations on $8^{3}\times4$ lattices, corresponding to the deconfined phase at a series of different values of $\thetaB$ and $\thetaI$.
A standard hybrid Monte Carlo simulation was observed to freeze in incorrect Polyakov loop sectors (i.e., in the one corresponding to a random initial configuration) even in this small volume.
To circumvent this problem, we included an additional update step after every twentieth trajectory, offering the system a $\mathrm{Z}(3)$ transformation in a random direction, $U_0\to U_0\cdot \exp(\pm \imath 2\pi/3)$ on a random time slice.
The action difference was computed exactly via the propagator matrix representation~\cite{TOUSSAINT1990248}, which allows to express the fermion determinant analytically as a function of the quark chemical potential.
Since the $\mathrm{Z}(3)$ transformation can also be viewed as a change of $\thetaB$ by $\pm 2\pi/3$, a single calculation of the eigenvalues of the propagator matrix suffices to get the up and down quark determinant both before and after the $\mathrm{Z}(3)$ rotation.

\begin{figure}[h]
    \centering
     \mbox{
     \subfigure[]{%
	\label{fig:updatesa}\includegraphics[height=6.cm]{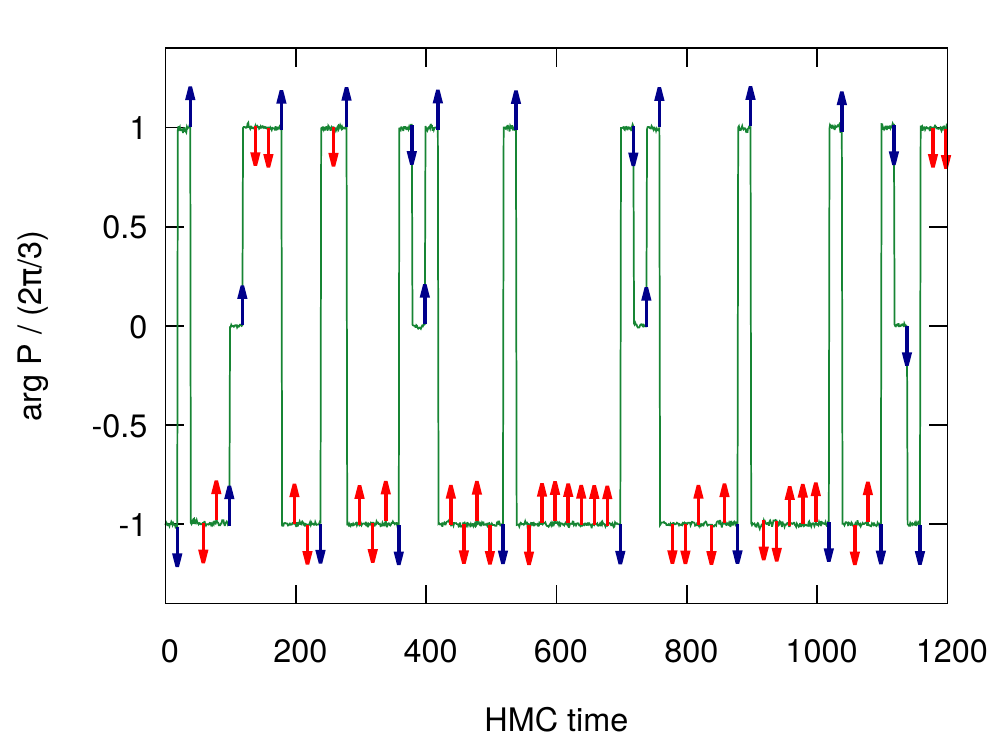}} \qquad
     \subfigure[]{%
	\label{fig:updatesb}\includegraphics[height=6.cm]{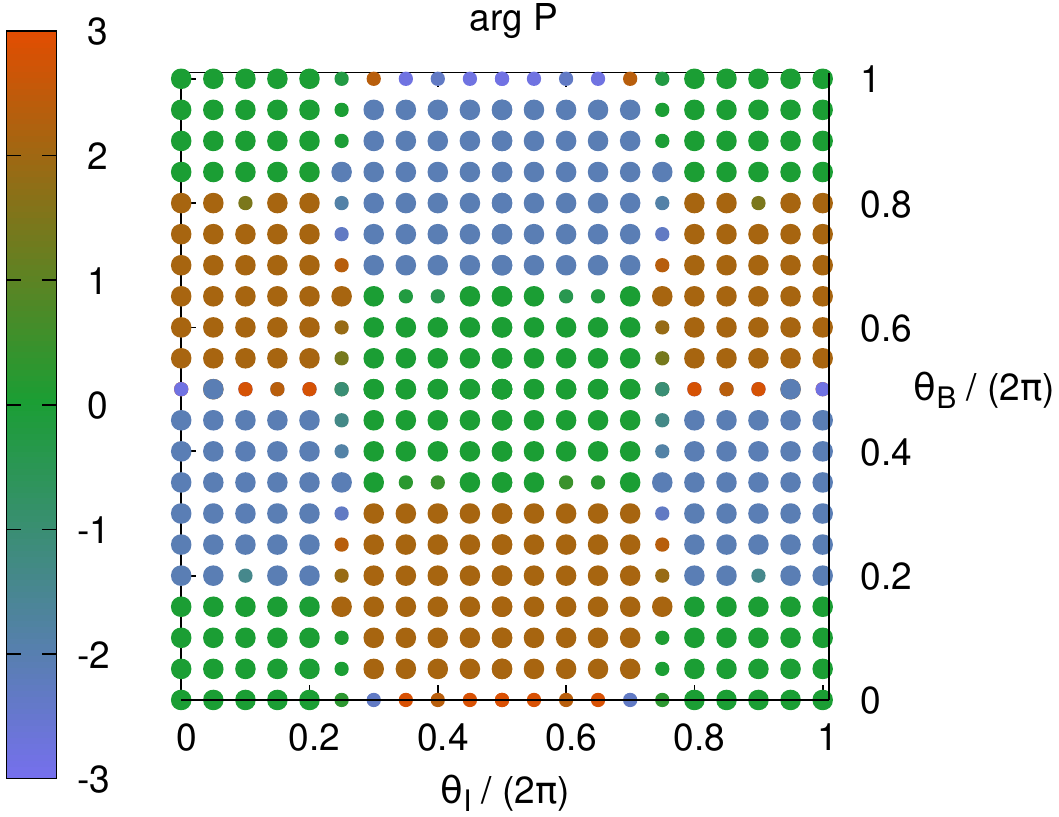}}
     }
     \caption{Left: Monte Carlo history of the phase of the Polyakov loop. After every twentieth trajectory a $\mathrm{Z}(3)$ update is proposed either in the clockwise (downward arrow) or counter-clockwise (upward arrow) direction in the complex plane. Accepted (rejected) proposals are indicated by blue (red) arrows. Right: Polyakov loop sectors in the plane of imaginary isospin and baryon chemical potentials, as obtained from lattice simulations. The color coding represent the phase of the Polyakov loop and the size of the dots is related to its variance (small dots indicate large variances).}
\end{figure}

The efficiency of the update procedure is visualized in~\cref{fig:updatesa}, which shows a section of the Monte Carlo history of the phase of $P$ at $\thetaB/(2\pi)=0.3$ and $\thetaI/(2\pi)=0.25$.
At these values the system is close to a $\mathrm{Z}(3)$ symmetric point and therefore frequent jumps between all three sectors are observed.
The average phase of the Polyakov loop, as measured using the simulations including these kind of updates, is shown in~\cref{fig:updatesb}.
The color coding is the same as in~\cref{fig:1a}.
For chemical potentials in the interior of the hexagons, the system selects one sector during thermalization and then also remains there for most of the simulation time. 
These lattice results are shown by the big dots. In turn, the small dots indicate runs, where frequent jumps between different sectors are observed -- this happens close to the boundaries between sectors, where $\mathrm{Z}(2)$ or $\mathrm{Z}(3)$ symmetry breaking occurs.
We remark that we only calculated the lower left quadrant of the plot and obtained the remaining parts by symmetry. 
In summary, the lattice results fully confirm the findings of the perturbative analysis.

%% file: AppendixSimulations.tex
\section{Overview of lattice ensembles} \label{appendix:simulations}

At each value of the quark mass $m$, spatial volume $\Nspat$, and $\beta$, four independent Markov chains were produced.
Typically, a single chain was thermalized starting from a thermalized configuration at a nearby value of $\beta$ at the same spatial volume and quark mass.
Taking into account autocorrelations~\cite{WOLFF2004143}, each chain was required to have at least $O(100)$ independent events for $B_4(\operatorname{Im} P)$, which is roughly the total number of configurations divided by twice the integrated autocorrelation time.
Most importantly, we used the number of standard deviations in the kurtosis values of the two maximally separated chains $n_\sigma^\text{max}(B_4)$ as an indicator for the reasonableness of the accumulated statistics.
This quantity should be as small as possible, and we typically aim at having it not much larger than 4.
A full overview of the ensembles used in this study is given in~\cref{tab:simulation_overview}.

\begin{table}[ht!]
    \setlength{\tabcolsep}{5mm}
    \newcommand{\CC}[1]{\cellcolor{#1}}
    \newcommand{\sep}{\,|\,}
    \renewcommand{\arraystretch}{1.25}
    \centering
    \begin{tabular}{@{}lS[table-format=1.3]*{4}{r}@{}}
        \toprule
        \multirow{2}{*}{$\Nf$} & {\multirow{2}{*}{$am$}} &  \multicolumn{4}{c}{Total statistics | Number of $\beta$ values | $n_\sigma^\text{max}(B_4)$ }\\
        &         &        \multicolumn{1}{c}{Aspect ratio $3$} & \multicolumn{1}{c}{Aspect ratio $4$} & \multicolumn{1}{c}{Aspect ratio $5$} & \multicolumn{1}{c}{Aspect ratio $6$} \\
        \midrule
        \multirow{3}{*}{2}
            & 0.04   &  1.2M\sep 3\sep 2.1 
            & 2.5M\sep 5\sep 3.3 
            & 3.1M\sep 5\sep 2.2 
            & 7.2M\sep 4\sep 2.9 
            \\
            & 0.05   & 1.2M\sep 3\sep 2.5 
            & 2.6M\sep 5\sep 2.5 
            & 7.4M\sep 5\sep 3.4 
            &
            \\
            & 0.06   &  1.6M\sep 3\sep 4.5 
            & 2.9M\sep 5\sep 2.4 
            & 6.6M\sep 6\sep 2.7
            &
            \\
        \bottomrule
    \end{tabular}
    \caption{%
      Table summarizing the ensembles used in this study. Due to large autocorrelation times, the number of configurations sharply increases as one goes to larger aspect ratios.
    }
    \label{tab:simulation_overview}
\end{table}

%% file: AppendixQuantitativeCollapse.tex
\section{Collapse quality optimization} \label{appendix:collapse_details}

Here we give further details regarding the quantitative collapse.
In order to construct the integrand of~\cref{eq:quality_collapse} as a continuous function of $x$, we use a cubic spline interpolation for each volume $\Nspat$ at a given quark mass $m$.
As it is costly to perform a dense scan in $\beta$ at each volume, one employs the standard technique of multi-histogram reweighting~\cite{FerrenbergSwendsenMH,newman1999monte}.
This gives us a dense training grid with $\Delta \beta \in [10^{-5}, 10^{-4}]$ and assures us that systematic errors associated with the interpolation are minimal.
Furthermore, for each ensemble we also generate $N_{\text{est}}=500$ estimators for $B_4$ using a bootstrap resampling of the raw data which is then used to produced reweighted samples.
Thus, this allows us to compute an analogous set of estimators $Q_{i,\text{est}}$ at each $\Delta x$ which will allow us to quantify the statistical error for $\{\betaC, \nu \}$.

\begin{figure} 
    \includegraphics[width=0.5\columnwidth]{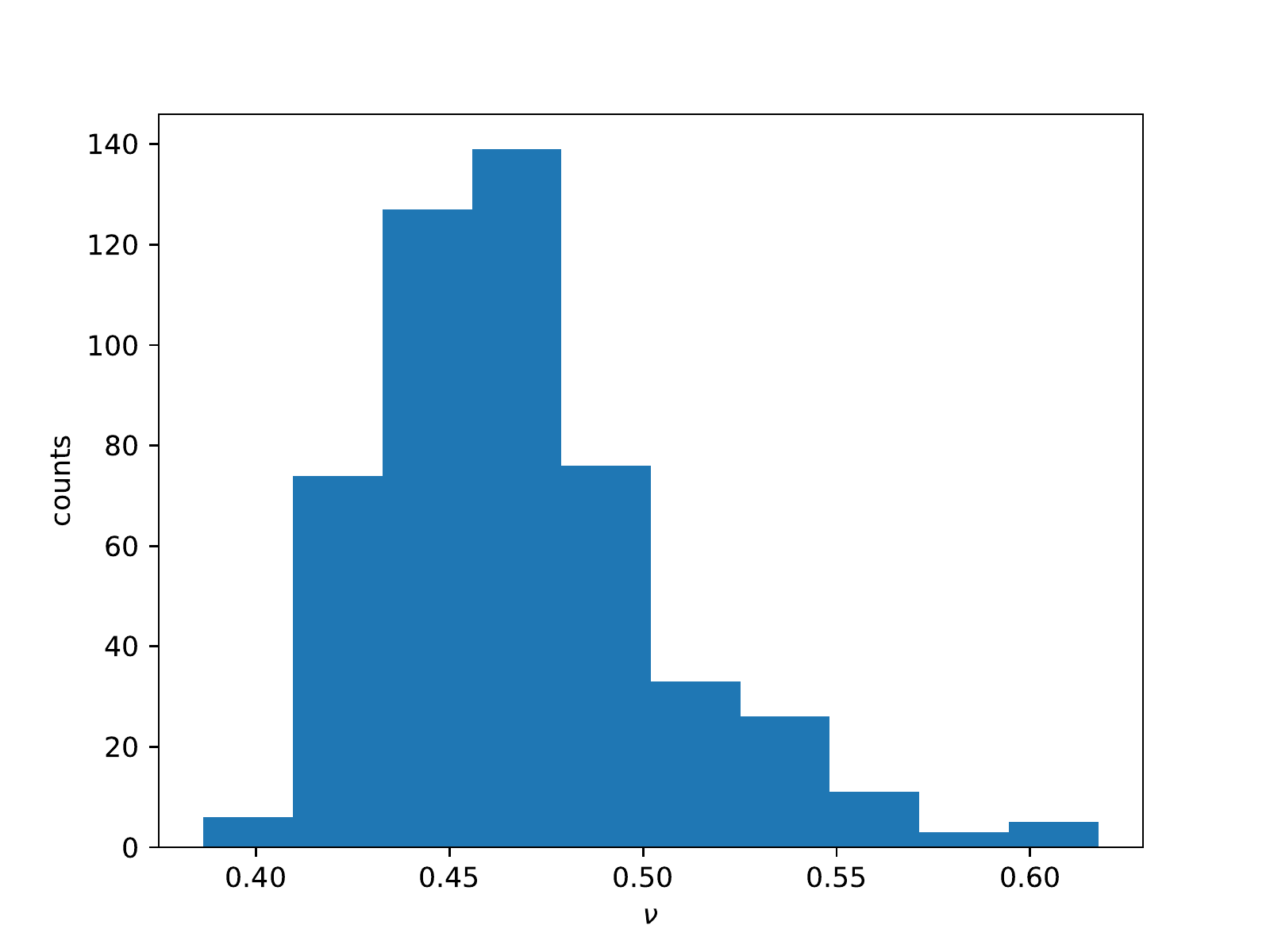}
    \caption{A histogram displaying the distribution of the critical exponent $\nu$ for the bootstrap estimators for $m=0.04$, $\Nspat=16,20$. A total of $N_{\text{boot}}=500$ were used with less than $100$ being discarded due to $Q$ not having a minimum in the allowed physical range.}
    \label{fig:nu_estimator_hist}
\end{figure}

\begin{figure*}
    \centering
    \subfigure[]{%
        \label{fig:Q_profile1}\includegraphics[width=0.5\columnwidth]{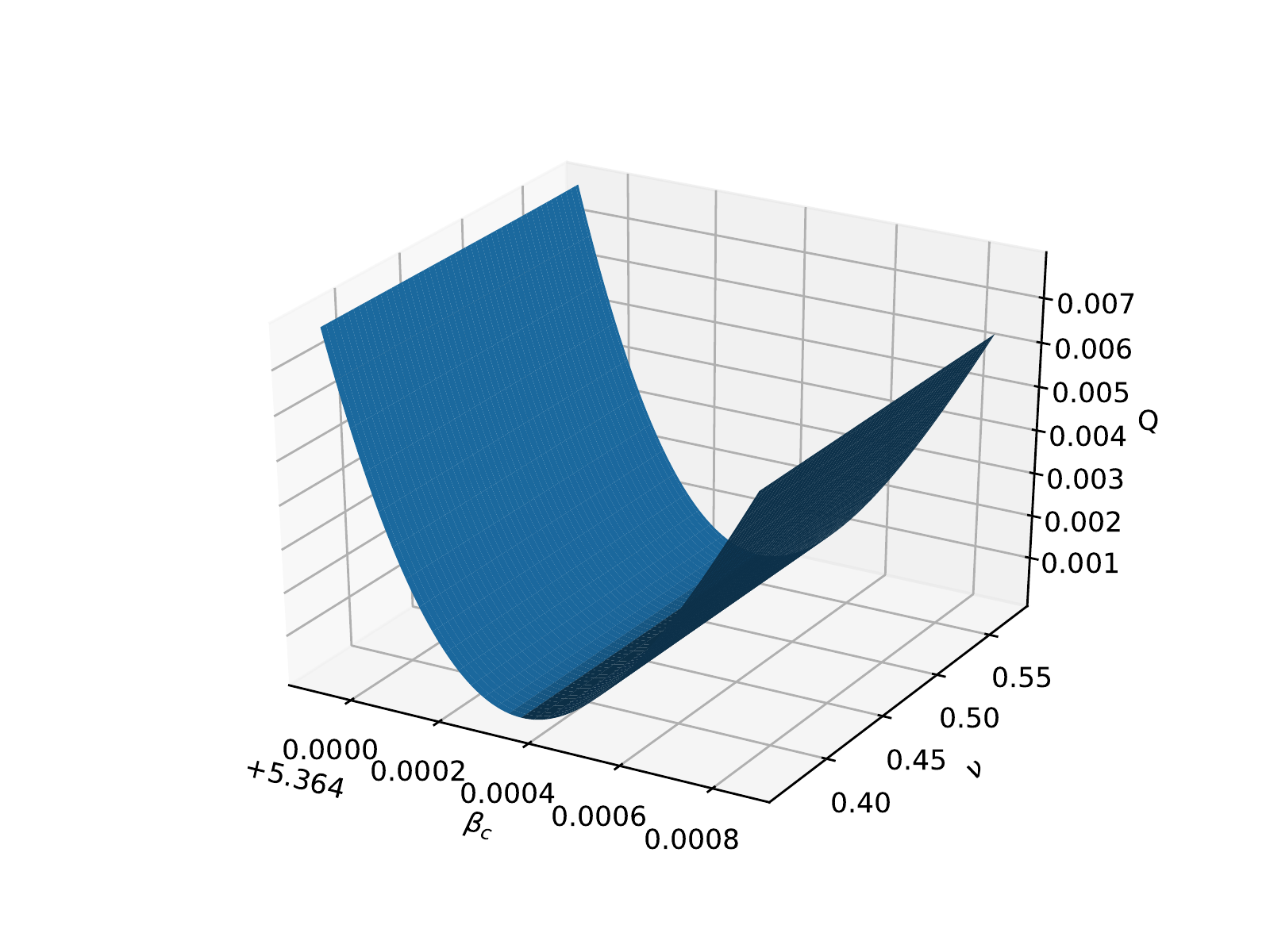}} \\
    \subfigure[]{%
        \label{fig:Q_profile2}\includegraphics[width=0.4\columnwidth]{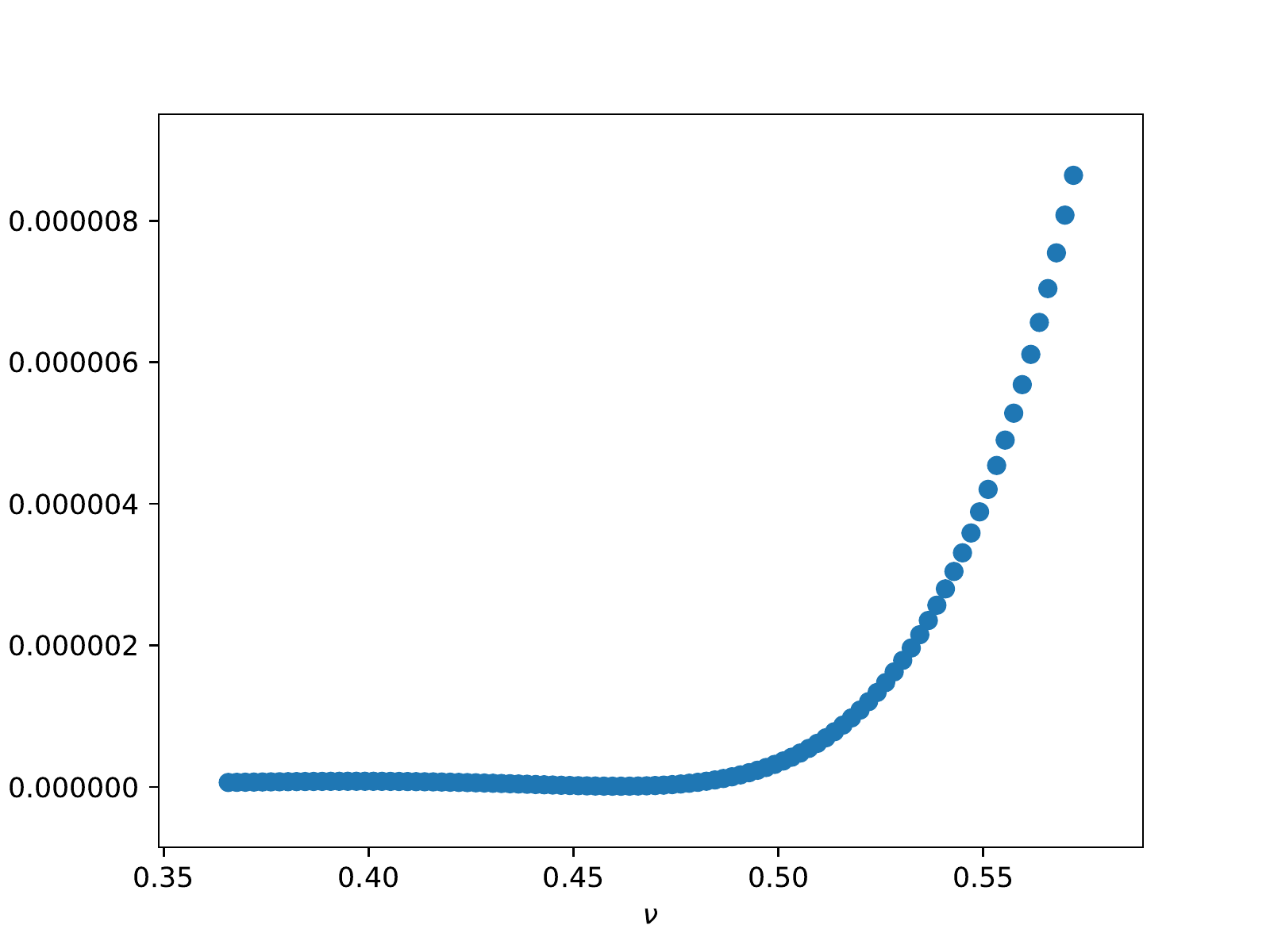}}
    \subfigure[]{%
        \label{fig:Q_profile3}\includegraphics[width=0.4\columnwidth]{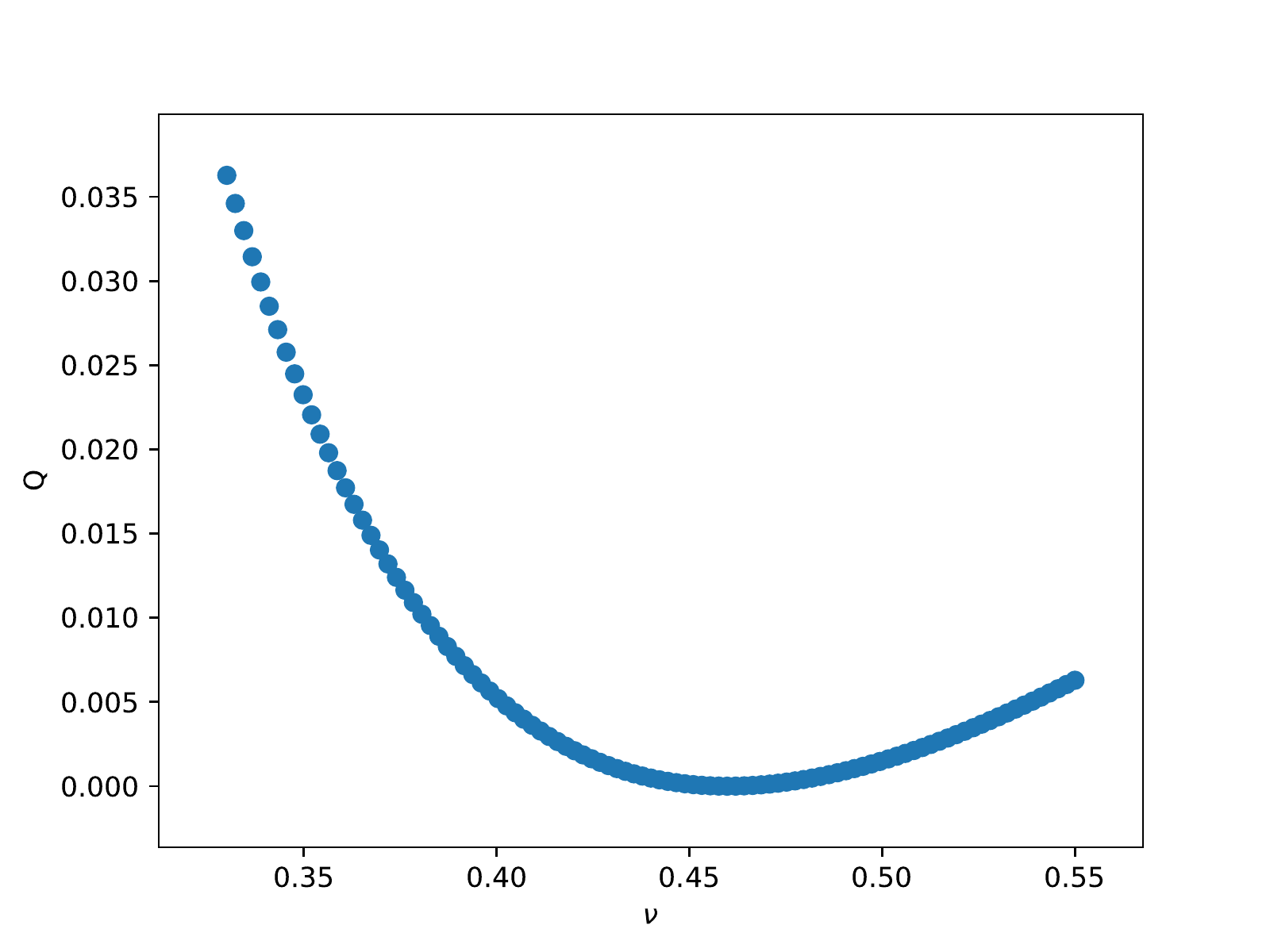}}
    \caption{\cref{fig:Q_profile1} displays the profile of $Q$ for $m=0.04$, $N_s=16,20$, and $\Delta x \approx 0.02$, in the region of the minimum. One immediately notices how steep $Q$ is with respect to $\betaC$ while the $\nu$ direction is very shallow. \cref{fig:Q_profile2} and \cref{fig:Q_profile3} present profiles of $Q$ and $\tilde{Q}$ as functions of $\nu$ only, for $\betaC = \bar{\betaC}$ for the same physical parameters as in \cref{fig:Q_profile1}. }. \label{fig:Q_profile}
\end{figure*}

The minimization of~\cref{eq:quality_collapse} and the estimation of the error for a given $\Delta x$ strongly depends on the search intervals in both critical parameters.
One must choose intervals
\beq 
\nu \in [\nu_{\text{min}}, \nu_{\text{max}}], ~
\betaC \in [\beta_{\text{c},\text{min}}, \beta_{\text{c},\text{max}}],
\eeq 
which contain a minimum of $Q$, $(\tilde{\nu}, \tilde{\betaC} )$, where $\partial Q/\partial \lambda_i|_{\lambda_i=\tilde{\lambda_i}} = 0 $, $\lambda_i \in \{\nu, \betaC \}, ~i=1,2$.
Typically, $Q$ has a very steep profile in $\betaC$ around $\tilde{\betaC}$.
Thus, one can first perform the collapse with some initial, broad interval for $\betaC$.
This interval can be chosen by hand based on the approximate crossing point of the reweighted data plotted as a function of $\beta$.
Assuming a reasonable interval for $\nu$ was given (in general, $\nu \in [1/3, 0.63]$), this typically leads to very accurate estimate $\tilde{\betaC}$, from which a much narrower interval in $\betaC$ can be used in a refined search.
This procedure for refining the search interval in $\betaC$ is beneficial in that it allows one to select a larger interval in the scaling variable $x$ over which we integrate.
This is due to the fact that the maximum allowed value of the scaling variable is given by
\beq 
\Delta x_{\text{max}} = \max_i \left\{ \left( \beta^{(i)}_{\text{max}} - \beta_{\text{c},\text{max}} \right) N_{\text{s},i}^{1/\nu_{\text{max}}}, \left( \beta_{\text{c},\text{min}} - \beta^{(i)}_{\text{min}} \right) N_{\text{s},i}^{1/\nu_{\text{max}}} \right\},
\eeq 
where each spatial volume $V_i = N_{\text{s},i}^3$ contains data for the kurtosis in the interval $[\beta^{(i)}_{\text{min}}, \beta^{(i)}_{\text{max}}]$.
Thus by narrowing the region in $\betaC$, one can still obtain a reasonably large interval in $x$ without overly constraining the search interval in $\nu$.
We note here that in practice, we usually take $5$ equally-spaced values of $\Delta x \in (0, \Delta x_{\text{max}} ]$.
Unlike in the $\betaC$ direction, $Q$ is very flat in the $\nu$ direction which can lead to difficulties when choosing the interval.
To illustrate this, we plot the profile of $Q(\betaC, \nu)$ for $m=0.04$ at fixed $\Delta x$ in~\cref{fig:Q_profile}.
One must, then, be sure to choose the interval such that  $\bar{\nu} \in [ \nu_{\text{min}}, \nu_{\text{max}}]$, as choosing $\nu_{\text{min}}$ too large can miss the desired minimum completely.
Furthermore, when choosing the interval for $\nu$ one must also take into account the minimization of $Q$ on the bootstrap estimators.
These minimizations on the estimators yield $\bar\nu_i$, $i=1,\dots,N_{\text{est}}$.
In fact, it can be that $\bar\nu$ lies within our search interval while a fraction of the $\bar\nu_i$ lie outside of the interval.
We have found that with good enough statistics, one should choose $[ \nu_{\text{min}}, \nu_{\text{max}}]$ such that less than $20\%$ of the estimates for $\nu$ at any given $\Delta x$ lie within the search interval.
For estimators whose minimum $\bar\nu_i$ lies outside of $[ \nu_{\text{min}}, \nu_{\text{max}}]$, we have discarded these from our error estimates.
A depiction of a distribution for $\nu$ obtained from the bootstrap estimators at a given $\Delta x$ at $m=0.04$ is shown in~\cref{fig:nu_estimator_hist}.
From this distribution, one can estimate the error which will then be used in the extrapolation of our estimate of the critical exponent to $\Delta x \to 0$.

To gain a better understanding of the procedure for minimizing $Q$, we take a closer look at its relation to the kurtosis at the various spatial volumes. 
Assuming that the kurtosis $B_{4,i}$, on each lattice size $V_i$ is a smooth function around the critical point we can write
\beq \label{eq:expansion_binder}
B_{4,i}(\betaC + x V_i^a) = 
    B_{4,i}(\betaC) + 
    x V_i^a B_{4,i}^{\prime}(\betaC) + 
    \frac{1}{2} x^2 V_i^{2 a} B_{4,i}^{\prime \prime}(\betaC) +
    \mathcal{O}(x^2),
\eeq 
where $a \equiv (3\nu)^{-1}$.
Inserting this expression into~\cref{eq:quality_collapse} and integrating with respect to $x$, we can obtain the following approximate expression for $Q$
\beq \label{eq:expansion_Q}
Q(\betaC, a) &= Q_0(\betaC) + \frac{\Delta x^2}{4} Q_2(\betaC, a) + 
\mathcal{O}(\Delta x^4),
\eeq 
where we have introduced
\beq \label{eq:expansion_Q_0}
Q_0(\betaC) &=& N_V \sum_{i=1}^{N_V} \left[ B_{4,i}(\betaC) \right]^2 - \left[\sum_{i=1}^{N_V} B_{4,i}(\betaC) \right]^2, \\ \label{eq:expansion_Q_02}
Q_{02}(\betaC, a) &=& N_V \sum_{i=1}^{N_V} B_{4,i}(\betaC) V_i^{2 a} B_{4,i}^{\prime \prime}(\betaC)  -
    \sum_{i=1}^{N_V} B_{4,i}(\betaC) \sum_{j=1}^{N_V} V_j^{2 a} B_{4,i}^{\prime \prime}(\betaC), \\ \label{eq:expansion_Q_11}
Q_{11}(\betaC, a) &=& N_V \sum_{i=1}^{N_V} \left[ V_i^{a} B_{4,i}^\prime(\betaC) \right]^2 - \left[\sum_{i=1}^{N_V} V_i^a B_{4,i}^\prime(\betaC) \right]^2,
\eeq
and have defined $Q_2 \equiv Q_{02} + Q_{11}$.
One immediately notices that~\cref{eq:expansion_Q_0} is independent of the critical exponent.
The minimization of $Q_0$ has a clear meaning: $Q_0$ is just the variance of $B_4$ over the various lattice sizes at a given value of $\betaC$.
In particular, if we indeed have a true intersection of all the cumulants, then at that point $Q_0 \to 0$, and thus this term achieves its minimum.
A further consequence if there is a true intersection point is that $Q_{02}$ also vanishes.
This is evident from~\cref{eq:expansion_Q_02}.
Finally, we note that the quantity $Q_{11}$ can be interpreted as the variance of the set of rescaled derivatives of the kurtosis and depends on both critical parameters. 

What the above discussion implies is that for small values of $\Delta x$, one can encounter problems determining the critical exponent $\nu$.
This scenario can frequently occur as it is not always possible to have data which cover a large enough range in the scaling variable $x$. 
One possibility to alleviate this problem is to modify the $Q$ function which one attempts to minimize.
One can note that Q is the sum of the $L_2$ norms of the pairwise differences of $B_{4,i}$ functions.
Changing the norm to the Sobolev norm $H^1$ adds a 
$\nu$-dependent leading-order term which yields
\beq \label{eq:modified_Q_ver1}
\bar{Q} = \frac{1}{2\Delta x} 
    \int_{-\Delta x}^{\Delta x} dx \left\{
    N_V \sum_{i=1}^{N_V} \left[B_{4,i}(x)\right]^2 
    - \left[ \sum_{i=1}^{N_V} B_{4,i}(x) \right]^2 + 
    N_V \sum_{i=1}^{N_V} \left[B^\prime_{4,i}(x)\right]^2 
    - \left[ \sum_{i=1}^{N_V} B^\prime_{4,i}(x) \right]^2
\right\}. 
\eeq 
However, it turns out that the minimization of this function still suffers from the problem that it is possible to return a value of $\nu$ which does not lie within the desired physical range.
The reason for this is that if we take $\Delta x$ to zero while assuming that we have fixed $\betaC$ to the exact intersection point, the minimization in $\nu$ is equivalent to finding the minimum of the variance of $B^\prime_{4,i}(x=0) = (1/N_{\text{s},i}^{1/\nu}) B^\prime_{4,i}(\beta=\betaC)$.
By decreasing $\nu$, it is possible to make all $B^\prime_{4,i}(x=0)$ less than any given positive value, so that the variance, and therefore $\bar{Q}$ be itself, would go to zero at $\nu \to 0$.
To remove this problem, we can normalize the two pairs of terms in~\cref{eq:modified_Q_ver1} by their corresponding mean values at $x=0$
\beq \label{eq:modified_Q_ver2}
\tilde{Q} = \frac{1}{2\Delta x} 
    \int_{-\Delta x}^{\Delta x} dx \left\{
    \frac{N_V \sum_{i=1}^{N_V} \left[B_{4,i}(x)\right]^2 
    - \left[ \sum_{i=1}^{N_V} B_{4,i}(x) \right]^2}{\left(\frac{1}{N_V} \sum_{i=1}^{N_V} B_{4,i}(0)\right)^2} + 
    \frac{N_V \sum_{i=1}^{N_V} \left[B^\prime_{4,i}(x)\right]^2 
    - \left[ \sum_{i=1}^{N_V} B^\prime_{4,i}(x) \right]^2}{\left(\frac{1}{N_V} \sum_{i=1}^{N_V} B^\prime_{4,i}(0)\right)^2}
\right\}. 
\eeq 
The profile of $\tilde{Q}$ defined in this way has a clear minimum as can be seen from~\cref{fig:Q_profile3}.

\begin{table}[t]
    \setlength{\tabcolsep}{5mm}
    \renewcommand{\arraystretch}{1.2}
    \begin{tabular}{ c  c  c  c  c  c  c }
    \toprule
      & 
    \multicolumn{3}{ c }{tuning the interval} &
    \multicolumn{3}{ c }{two-step procedure} 
    \\
    $m$ &
    $\beta_c$ & $\nu$ & $B_c$ &
    $\beta_c$ & $\nu$ & $B_c$ \\
    \midrule
     0.0400 & 
    5.36436(14) & 0.463(21) & 1.771(20) &
    5.36434(15) & 0.46(4) & 1.779(20) \\
     0.0500 & 
    5.37596(14) & 0.476(17) & 1.800(17) &
    5.37598(14) & 0.478(33) & 1.796(16) \\
     0.0600 & 
    5.38686(12) & 0.484(18) & 1.778(13) & 
    5.38687(13) & 0.49(4) & 1.772(13) \\
    \bottomrule
    \end{tabular}
    \caption{
    The final results and comparison of two minimisation procedures. 
    For each mass the critical coupling constant $\betaC$ and the critical index $\nu$ were obtained from 
    optimisation of the collapse for spatial sizes $N_s = 16, 20$. 
    The first set of columns corresponds to performing the minimisation explicitly tuning the range of $\betaC$ and $\nu$.
    The second set of columns describes the results of the two-step minimisation procedure described in this Appendix.
    }
\label{tbl:final_RW_Q_table}
\end{table}

To obtain our final estimates for the critical parameters of the theory, we have used the following strategy.
First, an initial scan is made using a continuous minimization algorithm for $\tilde{Q}$.
This gives us an initial estimate for both the critical coupling and exponent. Using this information, one can construct a smaller search interval in both variables taking for each parameter the interval of $\pm 3 \sigma$ around the previously obtained minimum.
Then a discrete second search for the minimum is performed for the original $Q$, where each parameter interval is divided into 100 subintervals.
An alternative continuous search for the minimum of the original $Q$ in the estimated small region was also attempted giving compatible parameter values.
The results are displayed in~\cref{tbl:final_RW_Q_table}. 
The critical parameters obtained from the initial search using $\tilde{Q}$ were found to have good agreement with the two subsequent searches with $Q$.
This gives one added confidence in the results of the procedure.

%% file: main.bbl
\begin{thebibliography}{45}%
\makeatletter
\providecommand \@ifxundefined [1]{%
 \@ifx{#1\undefined}
}%
\providecommand \@ifnum [1]{%
 \ifnum #1\expandafter \@firstoftwo
 \else \expandafter \@secondoftwo
 \fi
}%
\providecommand \@ifx [1]{%
 \ifx #1\expandafter \@firstoftwo
 \else \expandafter \@secondoftwo
 \fi
}%
\providecommand \natexlab [1]{#1}%
\providecommand \enquote  [1]{``#1''}%
\providecommand \bibnamefont  [1]{#1}%
\providecommand \bibfnamefont [1]{#1}%
\providecommand \citenamefont [1]{#1}%
\providecommand \href@noop [0]{\@secondoftwo}%
\providecommand \href [0]{\begingroup \@sanitize@url \@href}%
\providecommand \@href[1]{\@@startlink{#1}\@@href}%
\providecommand \@@href[1]{\endgroup#1\@@endlink}%
\providecommand \@sanitize@url [0]{\catcode `\\12\catcode `\$12\catcode
  `\&12\catcode `\#12\catcode `\^12\catcode `\_12\catcode `\%12\relax}%
\providecommand \@@startlink[1]{}%
\providecommand \@@endlink[0]{}%
\providecommand \url  [0]{\begingroup\@sanitize@url \@url }%
\providecommand \@url [1]{\endgroup\@href {#1}{\urlprefix }}%
\providecommand \urlprefix  [0]{URL }%
\providecommand \Eprint [0]{\href }%
\providecommand \doibase [0]{http://dx.doi.org/}%
\providecommand \selectlanguage [0]{\@gobble}%
\providecommand \bibinfo  [0]{\@secondoftwo}%
\providecommand \bibfield  [0]{\@secondoftwo}%
\providecommand \translation [1]{[#1]}%
\providecommand \BibitemOpen [0]{}%
\providecommand \bibitemStop [0]{}%
\providecommand \bibitemNoStop [0]{.\EOS\space}%
\providecommand \EOS [0]{\spacefactor3000\relax}%
\providecommand \BibitemShut  [1]{\csname bibitem#1\endcsname}%
\let\auto@bib@innerbib\@empty
\bibitem [{\citenamefont {de~Forcrand}\ and\ \citenamefont
  {Philipsen}(2002)}]{deForcrand:2002hgr}%
  \BibitemOpen
  \bibfield  {author} {\bibinfo {author} {\bibfnamefont {P.}~\bibnamefont
  {de~Forcrand}}\ and\ \bibinfo {author} {\bibfnamefont {O.}~\bibnamefont
  {Philipsen}},\ }\href {\doibase 10.1016/S0550-3213(02)00626-0} {\bibfield
  {journal} {\bibinfo  {journal} {Nucl. Phys. B}\ }\textbf {\bibinfo {volume}
  {642}},\ \bibinfo {pages} {290} (\bibinfo {year} {2002})},\ \Eprint
  {http://arxiv.org/abs/hep-lat/0205016} {arXiv:hep-lat/0205016} \BibitemShut
  {NoStop}%
\bibitem [{\citenamefont {D'Elia}\ and\ \citenamefont
  {Lombardo}(2003)}]{DElia:2002tig}%
  \BibitemOpen
  \bibfield  {author} {\bibinfo {author} {\bibfnamefont {M.}~\bibnamefont
  {D'Elia}}\ and\ \bibinfo {author} {\bibfnamefont {M.-P.}\ \bibnamefont
  {Lombardo}},\ }\href {\doibase 10.1103/PhysRevD.67.014505} {\bibfield
  {journal} {\bibinfo  {journal} {Phys. Rev. D}\ }\textbf {\bibinfo {volume}
  {67}},\ \bibinfo {pages} {014505} (\bibinfo {year} {2003})},\ \Eprint
  {http://arxiv.org/abs/hep-lat/0209146} {arXiv:hep-lat/0209146} \BibitemShut
  {NoStop}%
\bibitem [{\citenamefont {Roberge}\ and\ \citenamefont
  {Weiss}(1986)}]{ROBERGE1986734}%
  \BibitemOpen
  \bibfield  {author} {\bibinfo {author} {\bibfnamefont {A.}~\bibnamefont
  {Roberge}}\ and\ \bibinfo {author} {\bibfnamefont {N.}~\bibnamefont
  {Weiss}},\ }\href {\doibase https://doi.org/10.1016/0550-3213(86)90582-1}
  {\bibfield  {journal} {\bibinfo  {journal} {Nuclear Physics B}\ }\textbf
  {\bibinfo {volume} {275}},\ \bibinfo {pages} {734} (\bibinfo {year}
  {1986})}\BibitemShut {NoStop}%
\bibitem [{\citenamefont {de~Forcrand}\ and\ \citenamefont
  {Philipsen}(2003)}]{deForcrand:2003vyj}%
  \BibitemOpen
  \bibfield  {author} {\bibinfo {author} {\bibfnamefont {P.}~\bibnamefont
  {de~Forcrand}}\ and\ \bibinfo {author} {\bibfnamefont {O.}~\bibnamefont
  {Philipsen}},\ }\href {\doibase 10.1016/j.nuclphysb.2003.09.005} {\bibfield
  {journal} {\bibinfo  {journal} {Nucl. Phys. B}\ }\textbf {\bibinfo {volume}
  {673}},\ \bibinfo {pages} {170} (\bibinfo {year} {2003})},\ \Eprint
  {http://arxiv.org/abs/hep-lat/0307020} {arXiv:hep-lat/0307020} \BibitemShut
  {NoStop}%
\bibitem [{\citenamefont {de~Forcrand}\ and\ \citenamefont
  {Philipsen}(2007)}]{deForcrand:2006pv}%
  \BibitemOpen
  \bibfield  {author} {\bibinfo {author} {\bibfnamefont {P.}~\bibnamefont
  {de~Forcrand}}\ and\ \bibinfo {author} {\bibfnamefont {O.}~\bibnamefont
  {Philipsen}},\ }\href {\doibase 10.1088/1126-6708/2007/01/077} {\bibfield
  {journal} {\bibinfo  {journal} {JHEP}\ }\textbf {\bibinfo {volume} {01}},\
  \bibinfo {pages} {077} (\bibinfo {year} {2007})},\ \Eprint
  {http://arxiv.org/abs/hep-lat/0607017} {arXiv:hep-lat/0607017} \BibitemShut
  {NoStop}%
\bibitem [{\citenamefont {D'Elia}\ and\ \citenamefont
  {Sanfilippo}(2009)}]{DElia:2009bzj}%
  \BibitemOpen
  \bibfield  {author} {\bibinfo {author} {\bibfnamefont {M.}~\bibnamefont
  {D'Elia}}\ and\ \bibinfo {author} {\bibfnamefont {F.}~\bibnamefont
  {Sanfilippo}},\ }\href {\doibase 10.1103/PhysRevD.80.111501} {\bibfield
  {journal} {\bibinfo  {journal} {Phys. Rev. D}\ }\textbf {\bibinfo {volume}
  {80}},\ \bibinfo {pages} {111501} (\bibinfo {year} {2009})},\ \Eprint
  {http://arxiv.org/abs/0909.0254} {arXiv:0909.0254 [hep-lat]} \BibitemShut
  {NoStop}%
\bibitem [{\citenamefont {Cea}\ \emph {et~al.}(2009)\citenamefont {Cea},
  \citenamefont {Cosmai}, \citenamefont {D'Elia}, \citenamefont {Manneschi},\
  and\ \citenamefont {Papa}}]{Cea:2009ba}%
  \BibitemOpen
  \bibfield  {author} {\bibinfo {author} {\bibfnamefont {P.}~\bibnamefont
  {Cea}}, \bibinfo {author} {\bibfnamefont {L.}~\bibnamefont {Cosmai}},
  \bibinfo {author} {\bibfnamefont {M.}~\bibnamefont {D'Elia}}, \bibinfo
  {author} {\bibfnamefont {C.}~\bibnamefont {Manneschi}}, \ and\ \bibinfo
  {author} {\bibfnamefont {A.}~\bibnamefont {Papa}},\ }\href {\doibase
  10.1103/PhysRevD.80.034501} {\bibfield  {journal} {\bibinfo  {journal} {Phys.
  Rev. D}\ }\textbf {\bibinfo {volume} {80}},\ \bibinfo {pages} {034501}
  (\bibinfo {year} {2009})},\ \Eprint {http://arxiv.org/abs/0905.1292}
  {arXiv:0905.1292 [hep-lat]} \BibitemShut {NoStop}%
\bibitem [{\citenamefont {de~Forcrand}\ and\ \citenamefont
  {Philipsen}(2010)}]{deForcrand:2010he}%
  \BibitemOpen
  \bibfield  {author} {\bibinfo {author} {\bibfnamefont {P.}~\bibnamefont
  {de~Forcrand}}\ and\ \bibinfo {author} {\bibfnamefont {O.}~\bibnamefont
  {Philipsen}},\ }\href {\doibase 10.1103/PhysRevLett.105.152001} {\bibfield
  {journal} {\bibinfo  {journal} {Phys. Rev. Lett.}\ }\textbf {\bibinfo
  {volume} {105}},\ \bibinfo {pages} {152001} (\bibinfo {year} {2010})},\
  \Eprint {http://arxiv.org/abs/1004.3144} {arXiv:1004.3144 [hep-lat]}
  \BibitemShut {NoStop}%
\bibitem [{\citenamefont {Bonati}\ \emph {et~al.}(2011)\citenamefont {Bonati},
  \citenamefont {Cossu}, \citenamefont {D'Elia},\ and\ \citenamefont
  {Sanfilippo}}]{Bonati:2010gi}%
  \BibitemOpen
  \bibfield  {author} {\bibinfo {author} {\bibfnamefont {C.}~\bibnamefont
  {Bonati}}, \bibinfo {author} {\bibfnamefont {G.}~\bibnamefont {Cossu}},
  \bibinfo {author} {\bibfnamefont {M.}~\bibnamefont {D'Elia}}, \ and\ \bibinfo
  {author} {\bibfnamefont {F.}~\bibnamefont {Sanfilippo}},\ }\href {\doibase
  10.1103/PhysRevD.83.054505} {\bibfield  {journal} {\bibinfo  {journal} {Phys.
  Rev. D}\ }\textbf {\bibinfo {volume} {83}},\ \bibinfo {pages} {054505}
  (\bibinfo {year} {2011})},\ \Eprint {http://arxiv.org/abs/1011.4515}
  {arXiv:1011.4515 [hep-lat]} \BibitemShut {NoStop}%
\bibitem [{\citenamefont {Bonati}\ \emph {et~al.}(2014)\citenamefont {Bonati},
  \citenamefont {de~Forcrand}, \citenamefont {D'Elia}, \citenamefont
  {Philipsen},\ and\ \citenamefont {Sanfilippo}}]{Bonati:2014kpa}%
  \BibitemOpen
  \bibfield  {author} {\bibinfo {author} {\bibfnamefont {C.}~\bibnamefont
  {Bonati}}, \bibinfo {author} {\bibfnamefont {P.}~\bibnamefont {de~Forcrand}},
  \bibinfo {author} {\bibfnamefont {M.}~\bibnamefont {D'Elia}}, \bibinfo
  {author} {\bibfnamefont {O.}~\bibnamefont {Philipsen}}, \ and\ \bibinfo
  {author} {\bibfnamefont {F.}~\bibnamefont {Sanfilippo}},\ }\href {\doibase
  10.1103/PhysRevD.90.074030} {\bibfield  {journal} {\bibinfo  {journal} {Phys.
  Rev. D}\ }\textbf {\bibinfo {volume} {90}},\ \bibinfo {pages} {074030}
  (\bibinfo {year} {2014})},\ \Eprint {http://arxiv.org/abs/1408.5086}
  {arXiv:1408.5086 [hep-lat]} \BibitemShut {NoStop}%
\bibitem [{\citenamefont {Philipsen}\ and\ \citenamefont
  {Pinke}(2014)}]{Philipsen:2014rpa}%
  \BibitemOpen
  \bibfield  {author} {\bibinfo {author} {\bibfnamefont {O.}~\bibnamefont
  {Philipsen}}\ and\ \bibinfo {author} {\bibfnamefont {C.}~\bibnamefont
  {Pinke}},\ }\href {\doibase 10.1103/PhysRevD.89.094504} {\bibfield  {journal}
  {\bibinfo  {journal} {Phys. Rev. D}\ }\textbf {\bibinfo {volume} {89}},\
  \bibinfo {pages} {094504} (\bibinfo {year} {2014})},\ \Eprint
  {http://arxiv.org/abs/1402.0838} {arXiv:1402.0838 [hep-lat]} \BibitemShut
  {NoStop}%
\bibitem [{\citenamefont {Czaban}\ \emph {et~al.}(2016)\citenamefont {Czaban},
  \citenamefont {Cuteri}, \citenamefont {Philipsen}, \citenamefont {Pinke},\
  and\ \citenamefont {Sciarra}}]{Czaban:2015sas}%
  \BibitemOpen
  \bibfield  {author} {\bibinfo {author} {\bibfnamefont {C.}~\bibnamefont
  {Czaban}}, \bibinfo {author} {\bibfnamefont {F.}~\bibnamefont {Cuteri}},
  \bibinfo {author} {\bibfnamefont {O.}~\bibnamefont {Philipsen}}, \bibinfo
  {author} {\bibfnamefont {C.}~\bibnamefont {Pinke}}, \ and\ \bibinfo {author}
  {\bibfnamefont {A.}~\bibnamefont {Sciarra}},\ }\href {\doibase
  10.1103/PhysRevD.93.054507} {\bibfield  {journal} {\bibinfo  {journal} {Phys.
  Rev. D}\ }\textbf {\bibinfo {volume} {93}},\ \bibinfo {pages} {054507}
  (\bibinfo {year} {2016})},\ \Eprint {http://arxiv.org/abs/1512.07180}
  {arXiv:1512.07180 [hep-lat]} \BibitemShut {NoStop}%
\bibitem [{\citenamefont {Philipsen}\ and\ \citenamefont
  {Pinke}(2016)}]{Philipsen:2016hkv}%
  \BibitemOpen
  \bibfield  {author} {\bibinfo {author} {\bibfnamefont {O.}~\bibnamefont
  {Philipsen}}\ and\ \bibinfo {author} {\bibfnamefont {C.}~\bibnamefont
  {Pinke}},\ }\href {\doibase 10.1103/PhysRevD.93.114507} {\bibfield  {journal}
  {\bibinfo  {journal} {Phys. Rev. D}\ }\textbf {\bibinfo {volume} {93}},\
  \bibinfo {pages} {114507} (\bibinfo {year} {2016})},\ \Eprint
  {http://arxiv.org/abs/1602.06129} {arXiv:1602.06129 [hep-lat]} \BibitemShut
  {NoStop}%
\bibitem [{\citenamefont {Bonati}\ \emph {et~al.}(2016)\citenamefont {Bonati},
  \citenamefont {D'Elia}, \citenamefont {Mariti}, \citenamefont {Mesiti},
  \citenamefont {Negro},\ and\ \citenamefont {Sanfilippo}}]{Bonati:2016pwz}%
  \BibitemOpen
  \bibfield  {author} {\bibinfo {author} {\bibfnamefont {C.}~\bibnamefont
  {Bonati}}, \bibinfo {author} {\bibfnamefont {M.}~\bibnamefont {D'Elia}},
  \bibinfo {author} {\bibfnamefont {M.}~\bibnamefont {Mariti}}, \bibinfo
  {author} {\bibfnamefont {M.}~\bibnamefont {Mesiti}}, \bibinfo {author}
  {\bibfnamefont {F.}~\bibnamefont {Negro}}, \ and\ \bibinfo {author}
  {\bibfnamefont {F.}~\bibnamefont {Sanfilippo}},\ }\href {\doibase
  10.1103/PhysRevD.93.074504} {\bibfield  {journal} {\bibinfo  {journal} {Phys.
  Rev. D}\ }\textbf {\bibinfo {volume} {93}},\ \bibinfo {pages} {074504}
  (\bibinfo {year} {2016})},\ \Eprint {http://arxiv.org/abs/1602.01426}
  {arXiv:1602.01426 [hep-lat]} \BibitemShut {NoStop}%
\bibitem [{\citenamefont {Bonati}\ \emph {et~al.}(2019)\citenamefont {Bonati},
  \citenamefont {Calore}, \citenamefont {D'Elia}, \citenamefont {Mesiti},
  \citenamefont {Negro}, \citenamefont {Sanfilippo}, \citenamefont {Schifano},
  \citenamefont {Silvi},\ and\ \citenamefont {Tripiccione}}]{Bonati:2018fvg}%
  \BibitemOpen
  \bibfield  {author} {\bibinfo {author} {\bibfnamefont {C.}~\bibnamefont
  {Bonati}}, \bibinfo {author} {\bibfnamefont {E.}~\bibnamefont {Calore}},
  \bibinfo {author} {\bibfnamefont {M.}~\bibnamefont {D'Elia}}, \bibinfo
  {author} {\bibfnamefont {M.}~\bibnamefont {Mesiti}}, \bibinfo {author}
  {\bibfnamefont {F.}~\bibnamefont {Negro}}, \bibinfo {author} {\bibfnamefont
  {F.}~\bibnamefont {Sanfilippo}}, \bibinfo {author} {\bibfnamefont {S.~F.}\
  \bibnamefont {Schifano}}, \bibinfo {author} {\bibfnamefont {G.}~\bibnamefont
  {Silvi}}, \ and\ \bibinfo {author} {\bibfnamefont {R.}~\bibnamefont
  {Tripiccione}},\ }\href {\doibase 10.1103/PhysRevD.99.014502} {\bibfield
  {journal} {\bibinfo  {journal} {Phys. Rev. D}\ }\textbf {\bibinfo {volume}
  {99}},\ \bibinfo {pages} {014502} (\bibinfo {year} {2019})},\ \Eprint
  {http://arxiv.org/abs/1807.02106} {arXiv:1807.02106 [hep-lat]} \BibitemShut
  {NoStop}%
\bibitem [{\citenamefont {Philipsen}\ and\ \citenamefont
  {Sciarra}(2020{\natexlab{a}})}]{Philipsen:2019ouy}%
  \BibitemOpen
  \bibfield  {author} {\bibinfo {author} {\bibfnamefont {O.}~\bibnamefont
  {Philipsen}}\ and\ \bibinfo {author} {\bibfnamefont {A.}~\bibnamefont
  {Sciarra}},\ }\href {\doibase 10.1103/PhysRevD.101.014502} {\bibfield
  {journal} {\bibinfo  {journal} {Phys. Rev. D}\ }\textbf {\bibinfo {volume}
  {101}},\ \bibinfo {pages} {014502} (\bibinfo {year} {2020}{\natexlab{a}})},\
  \Eprint {http://arxiv.org/abs/1909.12253} {arXiv:1909.12253 [hep-lat]}
  \BibitemShut {NoStop}%
\bibitem [{\citenamefont {Cuteri}\ \emph {et~al.}(2022)\citenamefont {Cuteri},
  \citenamefont {Goswami}, \citenamefont {Karsch}, \citenamefont {Lahiri},
  \citenamefont {Neumann}, \citenamefont {Philipsen}, \citenamefont {Schmidt},\
  and\ \citenamefont {Sciarra}}]{Cuteri:2022vwk}%
  \BibitemOpen
  \bibfield  {author} {\bibinfo {author} {\bibfnamefont {F.}~\bibnamefont
  {Cuteri}}, \bibinfo {author} {\bibfnamefont {J.}~\bibnamefont {Goswami}},
  \bibinfo {author} {\bibfnamefont {F.}~\bibnamefont {Karsch}}, \bibinfo
  {author} {\bibfnamefont {A.}~\bibnamefont {Lahiri}}, \bibinfo {author}
  {\bibfnamefont {M.}~\bibnamefont {Neumann}}, \bibinfo {author} {\bibfnamefont
  {O.}~\bibnamefont {Philipsen}}, \bibinfo {author} {\bibfnamefont
  {C.}~\bibnamefont {Schmidt}}, \ and\ \bibinfo {author} {\bibfnamefont
  {A.}~\bibnamefont {Sciarra}},\ }\href@noop {} {\  (\bibinfo {year} {2022})},\
  \Eprint {http://arxiv.org/abs/2205.12707} {arXiv:2205.12707 [hep-lat]}
  \BibitemShut {NoStop}%
\bibitem [{\citenamefont {Vovchenko}\ \emph {et~al.}(2021)\citenamefont
  {Vovchenko}, \citenamefont {Brandt}, \citenamefont {Cuteri}, \citenamefont
  {Endr\H{o}di}, \citenamefont {Hajkarim},\ and\ \citenamefont
  {Schaffner-Bielich}}]{Vovchenko:2020crk}%
  \BibitemOpen
  \bibfield  {author} {\bibinfo {author} {\bibfnamefont {V.}~\bibnamefont
  {Vovchenko}}, \bibinfo {author} {\bibfnamefont {B.~B.}\ \bibnamefont
  {Brandt}}, \bibinfo {author} {\bibfnamefont {F.}~\bibnamefont {Cuteri}},
  \bibinfo {author} {\bibfnamefont {G.}~\bibnamefont {Endr\H{o}di}}, \bibinfo
  {author} {\bibfnamefont {F.}~\bibnamefont {Hajkarim}}, \ and\ \bibinfo
  {author} {\bibfnamefont {J.}~\bibnamefont {Schaffner-Bielich}},\ }\href
  {\doibase 10.1103/PhysRevLett.126.012701} {\bibfield  {journal} {\bibinfo
  {journal} {Phys. Rev. Lett.}\ }\textbf {\bibinfo {volume} {126}},\ \bibinfo
  {pages} {012701} (\bibinfo {year} {2021})},\ \Eprint
  {http://arxiv.org/abs/2009.02309} {arXiv:2009.02309 [hep-ph]} \BibitemShut
  {NoStop}%
\bibitem [{\citenamefont {Middeldorf-Wygas}\ \emph {et~al.}(2022)\citenamefont
  {Middeldorf-Wygas}, \citenamefont {Oldengott}, \citenamefont {B\"odeker},\
  and\ \citenamefont {Schwarz}}]{Middeldorf-Wygas:2020glx}%
  \BibitemOpen
  \bibfield  {author} {\bibinfo {author} {\bibfnamefont {M.~M.}\ \bibnamefont
  {Middeldorf-Wygas}}, \bibinfo {author} {\bibfnamefont {I.~M.}\ \bibnamefont
  {Oldengott}}, \bibinfo {author} {\bibfnamefont {D.}~\bibnamefont
  {B\"odeker}}, \ and\ \bibinfo {author} {\bibfnamefont {D.~J.}\ \bibnamefont
  {Schwarz}},\ }\href {\doibase 10.1103/PhysRevD.105.123533} {\bibfield
  {journal} {\bibinfo  {journal} {Phys. Rev. D}\ }\textbf {\bibinfo {volume}
  {105}},\ \bibinfo {pages} {123533} (\bibinfo {year} {2022})},\ \Eprint
  {http://arxiv.org/abs/2009.00036} {arXiv:2009.00036 [hep-ph]} \BibitemShut
  {NoStop}%
\bibitem [{\citenamefont {Brandt}\ \emph {et~al.}(2018)\citenamefont {Brandt},
  \citenamefont {Endrodi},\ and\ \citenamefont
  {Schmalzbauer}}]{Brandt:2017oyy}%
  \BibitemOpen
  \bibfield  {author} {\bibinfo {author} {\bibfnamefont {B.~B.}\ \bibnamefont
  {Brandt}}, \bibinfo {author} {\bibfnamefont {G.}~\bibnamefont {Endrodi}}, \
  and\ \bibinfo {author} {\bibfnamefont {S.}~\bibnamefont {Schmalzbauer}},\
  }\href {\doibase 10.1103/PhysRevD.97.054514} {\bibfield  {journal} {\bibinfo
  {journal} {Phys. Rev. D}\ }\textbf {\bibinfo {volume} {97}},\ \bibinfo
  {pages} {054514} (\bibinfo {year} {2018})},\ \Eprint
  {http://arxiv.org/abs/1712.08190} {arXiv:1712.08190 [hep-lat]} \BibitemShut
  {NoStop}%
\bibitem [{\citenamefont {Brandt}\ and\ \citenamefont
  {Endr\H{o}di}(2019)}]{Brandt:2018omg}%
  \BibitemOpen
  \bibfield  {author} {\bibinfo {author} {\bibfnamefont {B.~B.}\ \bibnamefont
  {Brandt}}\ and\ \bibinfo {author} {\bibfnamefont {G.}~\bibnamefont
  {Endr\H{o}di}},\ }\href {\doibase 10.1103/PhysRevD.99.014518} {\bibfield
  {journal} {\bibinfo  {journal} {Phys. Rev. D}\ }\textbf {\bibinfo {volume}
  {99}},\ \bibinfo {pages} {014518} (\bibinfo {year} {2019})},\ \Eprint
  {http://arxiv.org/abs/1810.11045} {arXiv:1810.11045 [hep-lat]} \BibitemShut
  {NoStop}%
\bibitem [{\citenamefont {Cuteri}\ \emph {et~al.}(2021)\citenamefont {Cuteri},
  \citenamefont {Philipsen},\ and\ \citenamefont {Sciarra}}]{Cuteri:2021ikv}%
  \BibitemOpen
  \bibfield  {author} {\bibinfo {author} {\bibfnamefont {F.}~\bibnamefont
  {Cuteri}}, \bibinfo {author} {\bibfnamefont {O.}~\bibnamefont {Philipsen}}, \
  and\ \bibinfo {author} {\bibfnamefont {A.}~\bibnamefont {Sciarra}},\ }\href
  {\doibase 10.1007/JHEP11(2021)141} {\bibfield  {journal} {\bibinfo  {journal}
  {JHEP}\ }\textbf {\bibinfo {volume} {11}},\ \bibinfo {pages} {141} (\bibinfo
  {year} {2021})},\ \Eprint {http://arxiv.org/abs/2107.12739} {arXiv:2107.12739
  [hep-lat]} \BibitemShut {NoStop}%
\bibitem [{\citenamefont {Pisarski}\ and\ \citenamefont
  {Wilczek}(1984)}]{Pisarski:1983ms}%
  \BibitemOpen
  \bibfield  {author} {\bibinfo {author} {\bibfnamefont {R.~D.}\ \bibnamefont
  {Pisarski}}\ and\ \bibinfo {author} {\bibfnamefont {F.}~\bibnamefont
  {Wilczek}},\ }\href {\doibase 10.1103/PhysRevD.29.338} {\bibfield  {journal}
  {\bibinfo  {journal} {Phys. Rev. D}\ }\textbf {\bibinfo {volume} {29}},\
  \bibinfo {pages} {338} (\bibinfo {year} {1984})}\BibitemShut {NoStop}%
\bibitem [{\citenamefont {Bazavov}\ \emph {et~al.}(2017)\citenamefont
  {Bazavov}, \citenamefont {Ding}, \citenamefont {Hegde}, \citenamefont
  {Karsch}, \citenamefont {Laermann}, \citenamefont {Mukherjee}, \citenamefont
  {Petreczky},\ and\ \citenamefont {Schmidt}}]{Bazavov:2017xul}%
  \BibitemOpen
  \bibfield  {author} {\bibinfo {author} {\bibfnamefont {A.}~\bibnamefont
  {Bazavov}}, \bibinfo {author} {\bibfnamefont {H.~T.}\ \bibnamefont {Ding}},
  \bibinfo {author} {\bibfnamefont {P.}~\bibnamefont {Hegde}}, \bibinfo
  {author} {\bibfnamefont {F.}~\bibnamefont {Karsch}}, \bibinfo {author}
  {\bibfnamefont {E.}~\bibnamefont {Laermann}}, \bibinfo {author}
  {\bibfnamefont {S.}~\bibnamefont {Mukherjee}}, \bibinfo {author}
  {\bibfnamefont {P.}~\bibnamefont {Petreczky}}, \ and\ \bibinfo {author}
  {\bibfnamefont {C.}~\bibnamefont {Schmidt}},\ }\href {\doibase
  10.1103/PhysRevD.95.074505} {\bibfield  {journal} {\bibinfo  {journal} {Phys.
  Rev. D}\ }\textbf {\bibinfo {volume} {95}},\ \bibinfo {pages} {074505}
  (\bibinfo {year} {2017})},\ \Eprint {http://arxiv.org/abs/1701.03548}
  {arXiv:1701.03548 [hep-lat]} \BibitemShut {NoStop}%
\bibitem [{\citenamefont {Ding}\ \emph {et~al.}(2019)\citenamefont {Ding} \emph
  {et~al.}}]{HotQCD:2019xnw}%
  \BibitemOpen
  \bibfield  {author} {\bibinfo {author} {\bibfnamefont {H.~T.}\ \bibnamefont
  {Ding}} \emph {et~al.} (\bibinfo {collaboration} {HotQCD}),\ }\href {\doibase
  10.1103/PhysRevLett.123.062002} {\bibfield  {journal} {\bibinfo  {journal}
  {Phys. Rev. Lett.}\ }\textbf {\bibinfo {volume} {123}},\ \bibinfo {pages}
  {062002} (\bibinfo {year} {2019})},\ \Eprint
  {http://arxiv.org/abs/1903.04801} {arXiv:1903.04801 [hep-lat]} \BibitemShut
  {NoStop}%
\bibitem [{\citenamefont {Dini}\ \emph {et~al.}(2022)\citenamefont {Dini},
  \citenamefont {Hegde}, \citenamefont {Karsch}, \citenamefont {Lahiri},
  \citenamefont {Schmidt},\ and\ \citenamefont {Sharma}}]{Dini:2021hug}%
  \BibitemOpen
  \bibfield  {author} {\bibinfo {author} {\bibfnamefont {L.}~\bibnamefont
  {Dini}}, \bibinfo {author} {\bibfnamefont {P.}~\bibnamefont {Hegde}},
  \bibinfo {author} {\bibfnamefont {F.}~\bibnamefont {Karsch}}, \bibinfo
  {author} {\bibfnamefont {A.}~\bibnamefont {Lahiri}}, \bibinfo {author}
  {\bibfnamefont {C.}~\bibnamefont {Schmidt}}, \ and\ \bibinfo {author}
  {\bibfnamefont {S.}~\bibnamefont {Sharma}},\ }\href {\doibase
  10.1103/PhysRevD.105.034510} {\bibfield  {journal} {\bibinfo  {journal}
  {Phys. Rev. D}\ }\textbf {\bibinfo {volume} {105}},\ \bibinfo {pages}
  {034510} (\bibinfo {year} {2022})},\ \Eprint
  {http://arxiv.org/abs/2111.12599} {arXiv:2111.12599 [hep-lat]} \BibitemShut
  {NoStop}%
\bibitem [{\citenamefont {Chabane}(2019)}]{AmineBachelorThesis}%
  \BibitemOpen
  \bibfield  {author} {\bibinfo {author} {\bibfnamefont {A.}~\bibnamefont
  {Chabane}},\ }\emph {\bibinfo {title} {QCD phase diagram with imaginary
  isospin chemical potential via perturbation theory}},\ \href@noop {}
  {\bibinfo {type} {Undergraduate honors thesis}},\ \bibinfo  {school} {Goethe
  Universität}, \bibinfo {address} {Frankfurt am Main} (\bibinfo {year}
  {2019})\BibitemShut {NoStop}%
\bibitem [{\citenamefont {Chabane}\ and\ \citenamefont
  {Endr\H{o}di}(2021)}]{Chabane:2021pfk}%
  \BibitemOpen
  \bibfield  {author} {\bibinfo {author} {\bibfnamefont {A.}~\bibnamefont
  {Chabane}}\ and\ \bibinfo {author} {\bibfnamefont {G.}~\bibnamefont
  {Endr\H{o}di}},\ }\href@noop {} {\bibfield  {journal} {\bibinfo  {journal}
  {PoS}\ }\textbf {\bibinfo {volume} {LATTICE2021}},\ \bibinfo {pages} {097}
  (\bibinfo {year} {2021})},\ \Eprint {http://arxiv.org/abs/2110.13536}
  {arXiv:2110.13536 [hep-lat]} \BibitemShut {NoStop}%
\bibitem [{\citenamefont {Follana}\ \emph {et~al.}(2007)\citenamefont
  {Follana}, \citenamefont {Mason}, \citenamefont {Davies}, \citenamefont
  {Hornbostel}, \citenamefont {Lepage}, \citenamefont {Shigemitsu},
  \citenamefont {Trottier},\ and\ \citenamefont {Wong}}]{HISQ}%
  \BibitemOpen
  \bibfield  {author} {\bibinfo {author} {\bibfnamefont {E.}~\bibnamefont
  {Follana}}, \bibinfo {author} {\bibfnamefont {Q.}~\bibnamefont {Mason}},
  \bibinfo {author} {\bibfnamefont {C.}~\bibnamefont {Davies}}, \bibinfo
  {author} {\bibfnamefont {K.}~\bibnamefont {Hornbostel}}, \bibinfo {author}
  {\bibfnamefont {G.~P.}\ \bibnamefont {Lepage}}, \bibinfo {author}
  {\bibfnamefont {J.}~\bibnamefont {Shigemitsu}}, \bibinfo {author}
  {\bibfnamefont {H.}~\bibnamefont {Trottier}}, \ and\ \bibinfo {author}
  {\bibfnamefont {K.}~\bibnamefont {Wong}},\ }\href {\doibase
  10.1103/PhysRevD.75.054502} {\bibfield  {journal} {\bibinfo  {journal} {Phys.
  Rev. D}\ }\textbf {\bibinfo {volume} {75}},\ \bibinfo {pages} {054502}
  (\bibinfo {year} {2007})}\BibitemShut {NoStop}%
\bibitem [{\citenamefont {Bazavov}\ \emph {et~al.}(2010)\citenamefont
  {Bazavov}, \citenamefont {Toussaint}, \citenamefont {Bernard}, \citenamefont
  {Laiho}, \citenamefont {DeTar}, \citenamefont {Levkova}, \citenamefont
  {Oktay}, \citenamefont {Gottlieb}, \citenamefont {Heller}, \citenamefont
  {Hetrick}, \citenamefont {Mackenzie}, \citenamefont {Sugar},\ and\
  \citenamefont {Van~de Water}}]{ASQTAD_review}%
  \BibitemOpen
  \bibfield  {author} {\bibinfo {author} {\bibfnamefont {A.}~\bibnamefont
  {Bazavov}}, \bibinfo {author} {\bibfnamefont {D.}~\bibnamefont {Toussaint}},
  \bibinfo {author} {\bibfnamefont {C.}~\bibnamefont {Bernard}}, \bibinfo
  {author} {\bibfnamefont {J.}~\bibnamefont {Laiho}}, \bibinfo {author}
  {\bibfnamefont {C.}~\bibnamefont {DeTar}}, \bibinfo {author} {\bibfnamefont
  {L.}~\bibnamefont {Levkova}}, \bibinfo {author} {\bibfnamefont {M.~B.}\
  \bibnamefont {Oktay}}, \bibinfo {author} {\bibfnamefont {S.}~\bibnamefont
  {Gottlieb}}, \bibinfo {author} {\bibfnamefont {U.~M.}\ \bibnamefont
  {Heller}}, \bibinfo {author} {\bibfnamefont {J.~E.}\ \bibnamefont {Hetrick}},
  \bibinfo {author} {\bibfnamefont {P.~B.}\ \bibnamefont {Mackenzie}}, \bibinfo
  {author} {\bibfnamefont {R.}~\bibnamefont {Sugar}}, \ and\ \bibinfo {author}
  {\bibfnamefont {R.~S.}\ \bibnamefont {Van~de Water}},\ }\href {\doibase
  10.1103/RevModPhys.82.1349} {\bibfield  {journal} {\bibinfo  {journal} {Rev.
  Mod. Phys.}\ }\textbf {\bibinfo {volume} {82}},\ \bibinfo {pages} {1349}
  (\bibinfo {year} {2010})}\BibitemShut {NoStop}%
\bibitem [{\citenamefont {Philipsen}\ and\ \citenamefont
  {Sciarra}(2020{\natexlab{b}})}]{PhilipsenSciarra2020}%
  \BibitemOpen
  \bibfield  {author} {\bibinfo {author} {\bibfnamefont {O.}~\bibnamefont
  {Philipsen}}\ and\ \bibinfo {author} {\bibfnamefont {A.}~\bibnamefont
  {Sciarra}},\ }\href {\doibase 10.1103/PhysRevD.101.014502} {\bibfield
  {journal} {\bibinfo  {journal} {Phys. Rev. D}\ }\textbf {\bibinfo {volume}
  {101}},\ \bibinfo {pages} {014502} (\bibinfo {year}
  {2020}{\natexlab{b}})}\BibitemShut {NoStop}%
\bibitem [{\citenamefont {Cardinali}\ \emph {et~al.}(2022)\citenamefont
  {Cardinali}, \citenamefont {D'Elia}, \citenamefont {Garosi},\ and\
  \citenamefont {Giordano}}]{Cardinali:2021fpu}%
  \BibitemOpen
  \bibfield  {author} {\bibinfo {author} {\bibfnamefont {M.}~\bibnamefont
  {Cardinali}}, \bibinfo {author} {\bibfnamefont {M.}~\bibnamefont {D'Elia}},
  \bibinfo {author} {\bibfnamefont {F.}~\bibnamefont {Garosi}}, \ and\ \bibinfo
  {author} {\bibfnamefont {M.}~\bibnamefont {Giordano}},\ }\href {\doibase
  10.1103/PhysRevD.105.014506} {\bibfield  {journal} {\bibinfo  {journal}
  {Phys. Rev. D}\ }\textbf {\bibinfo {volume} {105}},\ \bibinfo {pages}
  {014506} (\bibinfo {year} {2022})},\ \Eprint
  {http://arxiv.org/abs/2110.10029} {arXiv:2110.10029 [hep-lat]} \BibitemShut
  {NoStop}%
\bibitem [{\citenamefont {Fortunato}(2003)}]{Fortunato2003}%
  \BibitemOpen
  \bibfield  {author} {\bibinfo {author} {\bibfnamefont {S.}~\bibnamefont
  {Fortunato}},\ }\href {\doibase 10.1088/0305-4470/36/15/304} {\bibfield
  {journal} {\bibinfo  {journal} {J. Phys. A}\ }\textbf {\bibinfo {volume}
  {36}},\ \bibinfo {pages} {4269} (\bibinfo {year} {2003})}\BibitemShut
  {NoStop}%
\bibitem [{\citenamefont {Gattringer}(2010)}]{GATTRINGER2010179}%
  \BibitemOpen
  \bibfield  {author} {\bibinfo {author} {\bibfnamefont {C.}~\bibnamefont
  {Gattringer}},\ }\href {\doibase
  https://doi.org/10.1016/j.physletb.2010.05.013} {\bibfield  {journal}
  {\bibinfo  {journal} {Physics Letters B}\ }\textbf {\bibinfo {volume}
  {690}},\ \bibinfo {pages} {179} (\bibinfo {year} {2010})}\BibitemShut
  {NoStop}%
\bibitem [{\citenamefont {Endr\ifmmode~\mbox{\H{o}}\else \H{o}\fi{}di}\ \emph
  {et~al.}(2014)\citenamefont {Endr\ifmmode~\mbox{\H{o}}\else \H{o}\fi{}di},
  \citenamefont {Gattringer},\ and\ \citenamefont
  {Schadler}}]{EndrodiGattringer2014}%
  \BibitemOpen
  \bibfield  {author} {\bibinfo {author} {\bibfnamefont {G.}~\bibnamefont
  {Endr\ifmmode~\mbox{\H{o}}\else \H{o}\fi{}di}}, \bibinfo {author}
  {\bibfnamefont {C.}~\bibnamefont {Gattringer}}, \ and\ \bibinfo {author}
  {\bibfnamefont {H.-P.}\ \bibnamefont {Schadler}},\ }\href {\doibase
  10.1103/PhysRevD.89.054509} {\bibfield  {journal} {\bibinfo  {journal} {Phys.
  Rev. D}\ }\textbf {\bibinfo {volume} {89}},\ \bibinfo {pages} {054509}
  (\bibinfo {year} {2014})}\BibitemShut {NoStop}%
\bibitem [{\citenamefont {Ivanytskyi}\ \emph {et~al.}(2017)\citenamefont
  {Ivanytskyi}, \citenamefont {Bugaev}, \citenamefont {Nikonov}, \citenamefont
  {Ilgenfritz}, \citenamefont {Oliinychenko}, \citenamefont {Sagun},
  \citenamefont {Mishustin}, \citenamefont {Petrov},\ and\ \citenamefont
  {Zinovjev}}]{Ivanytskyi:2016wcp}%
  \BibitemOpen
  \bibfield  {author} {\bibinfo {author} {\bibfnamefont {A.~I.}\ \bibnamefont
  {Ivanytskyi}}, \bibinfo {author} {\bibfnamefont {K.~A.}\ \bibnamefont
  {Bugaev}}, \bibinfo {author} {\bibfnamefont {E.~G.}\ \bibnamefont {Nikonov}},
  \bibinfo {author} {\bibfnamefont {E.~M.}\ \bibnamefont {Ilgenfritz}},
  \bibinfo {author} {\bibfnamefont {D.~R.}\ \bibnamefont {Oliinychenko}},
  \bibinfo {author} {\bibfnamefont {V.~V.}\ \bibnamefont {Sagun}}, \bibinfo
  {author} {\bibfnamefont {I.~N.}\ \bibnamefont {Mishustin}}, \bibinfo {author}
  {\bibfnamefont {V.~K.}\ \bibnamefont {Petrov}}, \ and\ \bibinfo {author}
  {\bibfnamefont {G.~M.}\ \bibnamefont {Zinovjev}},\ }\href {\doibase
  10.1016/j.nuclphysa.2017.01.010} {\bibfield  {journal} {\bibinfo  {journal}
  {Nucl. Phys. A}\ }\textbf {\bibinfo {volume} {960}},\ \bibinfo {pages} {90}
  (\bibinfo {year} {2017})},\ \Eprint {http://arxiv.org/abs/1606.04710}
  {arXiv:1606.04710 [hep-lat]} \BibitemShut {NoStop}%
\bibitem [{\citenamefont {Bors\'anyi}\ \emph {et~al.}(2011)\citenamefont
  {Bors\'anyi}, \citenamefont {Danzer}, \citenamefont {Fodor}, \citenamefont
  {Gattringer},\ and\ \citenamefont {Schmidt}}]{Danzer2011}%
  \BibitemOpen
  \bibfield  {author} {\bibinfo {author} {\bibfnamefont {S.}~\bibnamefont
  {Bors\'anyi}}, \bibinfo {author} {\bibfnamefont {J.}~\bibnamefont {Danzer}},
  \bibinfo {author} {\bibfnamefont {Z.}~\bibnamefont {Fodor}}, \bibinfo
  {author} {\bibfnamefont {C.}~\bibnamefont {Gattringer}}, \ and\ \bibinfo
  {author} {\bibfnamefont {A.}~\bibnamefont {Schmidt}},\ }\href {\doibase
  10.1088/1742-6596/312/1/012005} {\bibfield  {journal} {\bibinfo  {journal}
  {J. Phys. Conf. Ser.}\ }\textbf {\bibinfo {volume} {312}},\ \bibinfo {pages}
  {012005} (\bibinfo {year} {2011})}\BibitemShut {NoStop}%
\bibitem [{\citenamefont {Fisher}(1967)}]{PhysicsPhysiqueFizika.3.255}%
  \BibitemOpen
  \bibfield  {author} {\bibinfo {author} {\bibfnamefont {M.~E.}\ \bibnamefont
  {Fisher}},\ }\href {\doibase 10.1103/PhysicsPhysiqueFizika.3.255} {\bibfield
  {journal} {\bibinfo  {journal} {Physics Physique Fizika}\ }\textbf {\bibinfo
  {volume} {3}},\ \bibinfo {pages} {255} (\bibinfo {year} {1967})}\BibitemShut
  {NoStop}%
\bibitem [{\citenamefont {Weiss}(1981)}]{Weiss:1980rj}%
  \BibitemOpen
  \bibfield  {author} {\bibinfo {author} {\bibfnamefont {N.}~\bibnamefont
  {Weiss}},\ }\href {\doibase 10.1103/PhysRevD.24.475} {\bibfield  {journal}
  {\bibinfo  {journal} {Phys. Rev. D}\ }\textbf {\bibinfo {volume} {24}},\
  \bibinfo {pages} {475} (\bibinfo {year} {1981})}\BibitemShut {NoStop}%
\bibitem [{\citenamefont {Weiss}(1982)}]{Weiss:1981ev}%
  \BibitemOpen
  \bibfield  {author} {\bibinfo {author} {\bibfnamefont {N.}~\bibnamefont
  {Weiss}},\ }\href {\doibase 10.1103/PhysRevD.25.2667} {\bibfield  {journal}
  {\bibinfo  {journal} {Phys. Rev. D}\ }\textbf {\bibinfo {volume} {25}},\
  \bibinfo {pages} {2667} (\bibinfo {year} {1982})}\BibitemShut {NoStop}%
\bibitem [{\citenamefont {Endr\H{o}di}(2014)}]{Endrodi:2014lja}%
  \BibitemOpen
  \bibfield  {author} {\bibinfo {author} {\bibfnamefont {G.}~\bibnamefont
  {Endr\H{o}di}},\ }\href {\doibase 10.1103/PhysRevD.90.094501} {\bibfield
  {journal} {\bibinfo  {journal} {Phys. Rev. D}\ }\textbf {\bibinfo {volume}
  {90}},\ \bibinfo {pages} {094501} (\bibinfo {year} {2014})},\ \Eprint
  {http://arxiv.org/abs/1407.1216} {arXiv:1407.1216 [hep-lat]} \BibitemShut
  {NoStop}%
\bibitem [{\citenamefont {Toussaint}(1990)}]{TOUSSAINT1990248}%
  \BibitemOpen
  \bibfield  {author} {\bibinfo {author} {\bibfnamefont {D.}~\bibnamefont
  {Toussaint}},\ }\href {\doibase https://doi.org/10.1016/0920-5632(90)90247-R}
  {\bibfield  {journal} {\bibinfo  {journal} {Nuclear Physics B - Proceedings
  Supplements}\ }\textbf {\bibinfo {volume} {17}},\ \bibinfo {pages} {248}
  (\bibinfo {year} {1990})}\BibitemShut {NoStop}%
\bibitem [{\citenamefont {Wolff}(2004)}]{WOLFF2004143}%
  \BibitemOpen
  \bibfield  {author} {\bibinfo {author} {\bibfnamefont {U.}~\bibnamefont
  {Wolff}},\ }\href {\doibase https://doi.org/10.1016/S0010-4655(03)00467-3}
  {\bibfield  {journal} {\bibinfo  {journal} {Computer Physics Communications}\
  }\textbf {\bibinfo {volume} {156}},\ \bibinfo {pages} {143} (\bibinfo {year}
  {2004})}\BibitemShut {NoStop}%
\bibitem [{\citenamefont {Ferrenberg}\ and\ \citenamefont
  {Swendsen}(1989)}]{FerrenbergSwendsenMH}%
  \BibitemOpen
  \bibfield  {author} {\bibinfo {author} {\bibfnamefont {A.~M.}\ \bibnamefont
  {Ferrenberg}}\ and\ \bibinfo {author} {\bibfnamefont {R.~H.}\ \bibnamefont
  {Swendsen}},\ }\href {\doibase 10.1103/PhysRevLett.63.1195} {\bibfield
  {journal} {\bibinfo  {journal} {Phys. Rev. Lett.}\ }\textbf {\bibinfo
  {volume} {63}},\ \bibinfo {pages} {1195} (\bibinfo {year}
  {1989})}\BibitemShut {NoStop}%
\bibitem [{\citenamefont {Newman}\ and\ \citenamefont
  {Barkema}(1999)}]{newman1999monte}%
  \BibitemOpen
  \bibfield  {author} {\bibinfo {author} {\bibfnamefont {M.}~\bibnamefont
  {Newman}}\ and\ \bibinfo {author} {\bibfnamefont {G.}~\bibnamefont
  {Barkema}},\ }\href {https://books.google.de/books?id=J5aLdDN4uFwC} {\emph
  {\bibinfo {title} {Monte Carlo Methods in Statistical Physics}}}\ (\bibinfo
  {publisher} {Clarendon Press},\ \bibinfo {year} {1999})\BibitemShut {NoStop}%
\end{thebibliography}%
